\documentclass{pasa}%

\usepackage{multirow}
\usepackage{graphicx}
\newcommand*{\spr}{{\sc Spritz}}
\newcommand*{\hers}{\textit{Herschel}\,}
\defcitealias{Bisigello2021}{B21}

\DeclareRobustCommand{\ion}[2]{\textup{#1\,\textsc{\lowercase{#2}}}}
\newcommand*\element[1][]{%
  \def\aa@element@tr{#1}%
  \aa@element
}

\title[{\it Simulating IR spectro-photometric surveys}]{Simulating infrared spectro-photometric surveys with a \spr}

\author[Bisigello,L. et al.]{Bisigello, L.$^1$\thanks{laura.bisigello@inaf.it}, Gruppioni, C.$^1$, Calura, F.$^1$, Feltre, A.$^1$, Pozzi, F.$^{1,2}$, Vignali, C.$^{2,1}$, Barchiesi L.$^{1,2}$, Rodighiero, G.$^{3,4}$, Negrello, M.$^{5}$, Carrera, F.J.$^{6}$, Dasyra, K.M.$^{7,8}$,Fernández-Ontiveros, J.A.$^{9,10}$, Giard, M.$^{11}$, Hatziminaoglou, E.$^{12}$, Kaneda, H.$^{13}$,
Lusso, E.$^{14,15}$, Pereira-Santaella, M.$^{16}$, P\'erez Gonz\'alez, P.G.$^{16}$, Ricci, C.$^{17,18}$, Schaerer, D.$^{19}$, Spinoglio, L.$^{9}$ and Wang, L.$^{20}$
\affil{$^1$INAF Osservatorio di Astrofisica e Scienza dello Spazio, via Gobetti 93/3, I-40129 Bologna, Italy}%
\affil{$^2$Dipartimento di Fisica e Astronomia, Università di Bologna, Via Gobetti 93/2, I-40129 Bologna, Italy}
\affil{$^3$Dipartimento di Fisica e Astronomia, Università di Padova, Vicolo dell'Osservatorio, 3, I-35122, Padova, Italy}
\affil{$^4$INAF Osservatorio Astronomico di Padova, vicolo dell'Osservatorio 5, I-35122 Padova, Italy}
\affil{$^5$School of Physics and Astronomy, Cardiff University, The Parade, Cardiff CF24 3AA, UK}
\affil{$^6$Instituto de Física de Cantabria (CSIC-U. Cantabria), Avenida de los Castros, 39005 Santander, Spain}
\affil{$^7$Dep. of Astrophysics, Astronomy \& Mechanics, Faculty of Physics, National and Kapodistrian University of Athens, Panepistimiopolis, Zografou 15784, Greece}
\affil{$^8$National Observatory of Athens, Institute for Astronomy, Astrophysics, Space Applications and Remote Sensing, Penteli 15236, Athens, Greece}
\affil{$^{9}$Istituto di Astrofisica e Planetologia Spaziali (INAF–IAPS), Via Fosso del Cavaliere 100, Roma I-00133, Italy}
\affil{$^{10}$Centro de Estudios de F\'{i}sica del Cosmos de Arag\'on, Unidad Asociada al CSIC, Plaza San Juan 1, E–44001 Teruel, Spain}
\affil{$^{11}$Institut de Recherche en Astrophysique et Plan\'etologie, Toulouse (CNRS-INSU), France}
\affil{$^{12}$European Southern Observatory, Karl-Schwarzschild-Str. 2, 85748 Garching bei M\"{u}nchen, Germany}
\affil{$^{13}$Graduate School of Science, Nagoya University,Furo-cho, Chikusa-ku, Nagoya 464-8602, Japan}
\affil{$^{14}$Dipartimento di Fisica e Astronomia, Università di Firenze, via G. Sansone 1, 50019 Sesto Fiorentino, Firenze, Italy}
\affil{$^{15}$ INAF – Osservatorio Astrofisico di Arcetri, 50125 Florence, Italy}
\affil{$^{16}$Centro de Astrobiolog\'ia (CSIC-INTA), Ctra. de Ajalvir, Km 4, 28850, Torrej\'on de Ardoz, Madrid, Spain}
\affil{$^{17}$N\'ucleo de Astronom\'ia de la Facultad de Ingenier\'ia, Universidad Diego Portales, Av. Ej\'ercito Libertador 441, Santiago 22, Chile}
\affil{$^{18}$Kavli Institute for Astronomy and Astrophysics, Peking University, Beijing 100871, People's Republic of China}
\affil{$^{19}$Observatoire de Gen\`eve, D\'epartement d'Astronomie, Universit\'e
de Gen\`eve, 51 Chemin Pegasi, 1290 Versoix, Switzerland}
\affil{$^{20}$SRON Netherlands Institute for Space Research, Landleven 12, 9747 AD, Groningen, The Netherlands}
}%

\jid{PASA}
\doi{10.1017/pas.\the\year.xxx}
\jyear{\the\year}

\usepackage{aas_macros}
\usepackage{hyperref}
\usepackage{xcolor}
\hypersetup{colorlinks,citecolor=blue,linkcolor=blue,urlcolor=blue}

\begin{document}

\begin{frontmatter}
\maketitle

\begin{abstract}

Mid- and far-infrared (IR) photometric and spectroscopic observations are fundamental to a full understanding of the dust-obscured Universe and the evolution of both star formation and black hole accretion in galaxies. In this work, using the specifications of the SPace Infrared telescope for Cosmology and Astrophysics (SPICA) as a baseline, we investigate the capability to study the dust-obscured Universe of mid- and far-IR photometry at 34 and 70 $\mu$m and low-resolution spectroscopy at 17-36 $\mu$m using the state-of-the-art Spectro-Photometric Realisations of Infrared-selected Targets at all-$z$ (\spr{}) simulation. This investigation is also compared to the expected performance of the Origins Space Telescope and the  Galaxy Evolution Probe. The photometric view of the Universe of a SPICA-like mission could cover not only bright objects (e.g. L$_{IR}>10^{12}\,L_{\odot}$) up to \textit{z}$=$10, but also normal galaxies (L$_{IR}<10^{11}\,L_{\odot}$) up to \textit{z}$\sim$4. At the same time, the spectroscopic observations of such mission could also allow us to estimate the redshifts and study the physical properties for thousands of star-forming galaxies and active galactic nuclei by observing the polycyclic aromatic hydrocarbons and a large set of IR nebular emission lines. In this way, a cold, 2.5-m size space telescope with spectro-photometric capability analogous to SPICA, could provide us with a complete three-dimensional (i.e. images and integrated spectra) view of the dust-obscured Universe and the physics governing galaxy evolution up to \textit{z}$\sim$4. 
\end{abstract}

\end{frontmatter}

\subsection*{Preface}
The articles of this special issue focus on some of the major scientific questions that a future IR observatory will be able to address. We adopt the SPace Infrared telescope for Cosmology and Astrophysics (SPICA) design as a baseline to demonstrate how to achieve the major scientific goals in the fields of galaxy evolution, Galactic star formation and protoplanetary disks formation and evolution. The studies developed for the SPICA mission serve as a reference for future work in the field, even though the mission proposal has been cancelled by ESA from its M5 competition. The mission concept of SPICA employs a 2.5 m telescope, actively cooled to below $\sim$8 K, and a suite of mid- to far-IR spectrometers and photometric cameras, equipped with state-of-the-art detectors \citep{roelfsema2018}. In particular, the SPICA Far-Infrared Instrument (SAFARI) is a grating spectrograph with low (R $\sim$ 200–300) and medium-resolution (R $\sim$ 3 000–11 000) observing modes instantaneously covering the 35–210 $\mu$m wavelength range. The SPICA Mid-Infrared Instrument (SMI) has three operating modes: a large field of view ($10' \times 12'$) low-resolution 17–36 $\mu$m imaging spectroscopic (R $\sim$ 50–120) mode and photometric camera at 34 $\mu$m (SMI-LR), a medium-resolution (R $\sim$ 1 300–2 300) grating spectrometer covering wavelengths of 18–36 $\mu$m (SMI-MR), and a high-resolution echelle module
(R $\sim$ 29 000) for the 10–18 $\mu$m domain (SMI-HR). Finally, BBOP, a large field of view ($2'.6 \times 2'.6$), three-channel (70, 200 and
350 $\mu$m) polarimetric camera complements the science payload.

\section{INTRODUCTION }
\label{sec:intro}
As new stars form in dusty environments, infrared (IR) observations are key to studying obscured star formation and, in turn, achieving a better understanding of galaxy formation and evolution. This is particularly important around the very active period known as the Cosmic Noon  \citep[i.e. \textit{z}$\sim$2;][]{Madau2014}, where dust absorbs and re-radiates in the IR, on average, 80$\%$ of the ultra-violet (UV) and optical radiation \citep{Reddy2012}. \par
At \textit{z}$>$3 our knowledge of the on-going star formation is mainly based on UV rest-frame observations \citep[e.g.][]{Bouwens2015,Livermore2017,Livermore2018,Oesch2018}, given the large availability of optical and near-IR observatories such as the Hubble Space Telescope ($HST$). UV observations have shown that dust corrections at \textit{z}$>$3 may be relatively small \citep[e.g.][]{Bouwens2016,Dunlop2017}. Conversely, studies based on emission lines have shown that UV-based star-formation rate may be underestimated for galaxies with high level of star formation, due to an increased importance of the dusty stellar populations in these galaxies \citep[e.g.][]{Reddy2015}. Similarly, recent observations of galaxies at \textit{z}$>$6 have shown the presence of a large quantity of dust \citep[e.g. $\sim10^6\,M_{\odot}$; ][]{Tamura2019}, which may be more concentrate and warm than clouds in local galaxies, increasing the UV attenuation \citep[e.g.][]{Sommovigo2020}\par
In addition, when studying the cosmic star-formation-rate density (SFRD) of the Universe, there are some tensions between the results obtained from UV observations \citep[e.g.][]{Schenker2013,Bouwens2015} and results based on IR or radio data \citep[e.g.][]{Magnelli2013,Novak2017,Gruppioni2020}, with the former indicating a steeper redshift evolution (i.e. lower SFRD at high-$z$) than the latter. This tension is at least partially due to UV observations not being representative of the whole galaxy population \citep{Bouwens2020}.  Indeed, recent observations have shown the non-negligible presence of a dusty and optically-dark galaxy population \citep[e.g.][]{Franco2018,Wang2019,Talia2021}, which contributes to 17$\%$ of the star-formation rate density at \textit{z}$\sim$5, as estimated by \citet{Gruppioni2020} from a sample of blindly detected far-IR galaxies in the ALMA Large Program to INvestigate survey \citep[ALPINE; ][]{LeFevre2020}. \par
Due to the capabilities of past and current IR and sub-millimetre observatories, e.g. \textit{Herschel}, \textit{Spitzer} and the Atacama Large Millimetre/submillimetre Array (ALMA), current observations have been limited to the brightest IR galaxies or to very small fields. A new IR telescope with both high sensitivity and large survey capability, such as a SPICA-like mission, the Origins Space Telescope\footnote{\url{http://origins.ipac.caltech.edu/}} \citep[OST; ][]{Leisawitz2019} or the Galaxy Evolution Probe \citep[GEP]{Glenn2018}, is therefore essential to improving our understanding of the dust-obscured Universe at different epochs. \par
One of the primary science goals of SPICA was the detection and analysis of the dust-obscured phases of galaxy evolution, including activity from both star formation and accretion onto super massive black holes. Combined low-resolution (LR) spectroscopy and photometry at IR wavelengths could enable the detection of polycyclic aromatic hydrocarbons (PAHs) features, fine-structure lines as well as the dust continuum emission at mid-IR wavelengths. At these wavelengths the light is not heavily affected by dust absorption, as happens instead at UV and optical wavelengths, and, therefore, it allows the analysis of the evolutionary phases which are difficult to study at shorter wavelengths. In addition, the advantage of the spectro-photometric observations at mid- to far-IR wavelengths is the possibility to analyse the physics behind the processes responsible for the observed galaxy evolution and the environment in which they take place.  \par
In this work we make use of the state-of-the-art Spectro-Photometric Realisations of Infrared-selected Targets at all-$z$ simulation \citep[\spr{}; ][ hereafter B21]{Bisigello2021} that is particularly suited for predictions of IR surveys and includes both star-forming galaxies and active galactic nuclei (AGN). Using this simulation and following the works by \citet{Kaneda2017}, \citet{Gruppioni2017} and \citet{Spinoglio2021}, we aim at exploring the capability of a SPICA-like cold 2.5 m telescope to drive significant progress in our knowledge on galaxy evolution. Given the dismissal of the SPICA mission concept from the M5 competition, we expanded this work to include a comparison with the OST and the GEP, two NASA concepts for cryogenic IR observatory with 5.9m and 2.0m primary mirrors, respectively. \par
This paper is organised as follows. Section \ref{sec:data_survey} introduces the \spr{} simulation, the SPICA spectro-photometric surveys used as references and the OST and GEP surveys. Section \ref{sec:pred} contains the different predictions for SPICA complemented, when possible, with OST and GEP expectations. We summarise our main findings in Section \ref{sec:conclusions}.
Throughout the paper, we consider a $\Lambda$CDM cosmology with $H_0=70\,{\rm km}\,{\rm s}^{-1}{\rm Mpc}^{-1} $, $\Omega_{\rm m}=0.27$ and $\Omega_\Lambda=0.73$.

%%%%%%%%%%%%%%%%%%%%%%%%%%%%%%%%%%%%%%%%%%%%%%%%%%%%%%%%%%%%%%%%%%%%%%%%%%%%%%%%%%%%%%%%%%%%%%%%%

\section{The \spr{} simulation and SPICA mock catalogues}\label{sec:data_survey}
\subsection{The \spr{} simulation}\label{sec:spritz}
To predict the galaxy population that could be observed by future IR missions,  we created mock observations using the newly developed \spr{} simulations \citepalias{Bisigello2021}.  All the mock catalogues considered in this paper are made publicly available\footnote{\label{mywebsite}\url{http://spritz.oas.inaf.it/}}. \par
Briefly, in \spr{} all simulated galaxies are extracted from a set of observed luminosity functions (LFs) or galaxy stellar mass functions (GSMF):
\begin{itemize}
    \item The \hers LFs measured by \citet{Gruppioni2013} for different galaxy populations such as normal star-forming galaxies (hereafter spirals), starburst (SB) and two composite systems with an AGN contribution that is only visible in the mid-IR and, even if with some obscuration, in the X-rays (SF-AGN and SB-AGN). The first composite AGN population, i.e. SF-AGN, is common at \textit{z}$=$1-2 and includes low-luminosity AGN, while the second one, i.e. SB-AGN, is common at \textit{z}$>$2, hosts a bright, but heavily obscured AGN and has a high SFR (i.e. log$_{10}(sSFR/yr)\sim$8 in the simulation). All these LFs are described by a modified Schechter function \citep{Saunders1990}, whose parameter values are summarised in Table 1 in \citetalias{Bisigello2021}. %: 
    % \begin{equation}
    % \begin{aligned} %& set the alignment
    % &\Phi(L)dlog_{10}L=\\
    % &\Phi^{*}\left(\frac{L}{L^{*}}\right)^{(1-\alpha)}exp\left[-\frac{1}{2\sigma^{2}}log^{2}_{10}\left(1+\frac{L}{L^{*}}\right)\right]dlog_{10}L
    % \end{aligned}
    % \end{equation}
    At \textit{z}$>$3 the LFs are extrapolated assuming a constant characteristic luminosity and a decreasing number density at the knee as $\propto(1+z)^{k_{\Phi}}$, with $k_{\Phi}=-$4 to $-$1. These values have been chosen to explore a large range of possible density evolution and are consistent with recent observations of the total IR LF observed at \textit{z}$>$3 \citep{Gruppioni2020}.
    \item The obscured and un-obscured AGN (AGN2 and AGN1, respectively) LF, as derived by \citetalias{Bisigello2021} using \hers observations along with a set of far-ultraviolet observations up to \textit{z}$=$5. The LF is described by a modified Schechter function and the evolution of the parameters describing the function at \textit{z}$<$5 is extrapolated at higher redshift \citepalias[see Table 1 in ][]{Bisigello2021}.
    \item The K-band LF of elliptical galaxies (Ell), taken from the average LF of \citet{Cirasuolo2007}, \citet{Arnouts2007} and \citet{Beare2019}. The LF is extrapolated for galaxies at \textit{z}$>$2 by keeping the characteristic luminosity constant and with a number density at the knee decreasing as $\propto(1+z)^{-1}$.
    \item The GSMF of dwarf irregular galaxies (Irr) by \citet{Huertas-Company2016}. For consistency with the \hers LFs, at \textit{z}$>$3 we extrapolated the GSMF, which is not well studied at these high-$z$, by keeping the characteristic mass constant and decreasing the number density at the knee as $\propto(1+z)^{k_{\Phi}}$, with $k_{\Phi}$ ranging from -4 to -1.
\end{itemize} \par
The different galaxy populations and the references for their LF or GSMF are reported in Table \ref{tab:LFref}.\par

Mock fluxes for different current and future facilities, including SPICA, are derived by convolving the throughput of each filter with an empirical galaxy template associated to each galaxy population. We considered 35 empirical templates taken from \citet{Polletta2007,Rieke2009,Gruppioni2010} and \citet{Bianchi2018}. We included these templates as they are the same used by \citet{Gruppioni2013} to derive the \hers IR LF, except for dwarf galaxies which were not observed in large numbers by \hers. \par
The main physical properties for each simulated galaxy, such as stellar mass, AGN fraction, gas metallicity and intrinsic luminosity at different wavelengths, are derived from the associated template or from a broad set of empirical relations \citepalias[see ][ and references therein]{Bisigello2021}. The simulation also includes mock SMI and OST spectra, obtained considering the same galaxy templates used for deriving the simulated photometric data. \par
We added optical and IR nebular line emission to these templates, considering a set of empirical and theoretical relations \citepalias[see ][ and references therein]{Bisigello2021}. All lines include star formation and, if present, the AGN contribution. The AGN component is limited to the contribution of the narrow-line gas-emitting regions. Therefore, as the contribution of the broad-line regions is missing, the fluxes of the permitted lines for the simulated un-obscured AGN (AGN1) should be considered as lower limits. Similarly, the numbers of AGN1 with a detection of permitted lines should be considered as lower limits.\par
The large-scale structure is included in the simulation using the algorithm by \citet{Soneira1978} to extract positions based on a two-point correlation function. In particular, we assumed the power-law approximation by \citet{Limber1953} for the angular two-point correlation function $w(\theta)=A_{w}\theta^{1-\gamma}$, with $\gamma$=1.7, as suggested by observations \citep[e.g. ][]{Wang2013}, and A$_{w}$ evolving with the stellar mass \citep{Wake2011,Hatfield2016} but not with redshift \citep{Bethermin2015,Schreiber2015}. We highlight that the assumptions made on the clustering have no impact on the results presented in this work, as this information in used, at the moment of writing, only to create mock images and we do not analyse source blending.
\par

As detailed in \citetalias{Bisigello2021}, \spr{} agrees well with a broad set of observations (LFs and number counts) at different wavelengths. In particular, the simulation is in agreement, within the simulation uncertainties, with the mid- and far-IR number counts ($\sim$24-600 $\mu$m) and with the recent IR LF observed by \citet{Gruppioni2020} up to \textit{z}$\sim$6, showing the reliability of the extrapolation at \textit{z}$>$3, at least at IR wavelengths. In addition, \citet{Gruppioni2020} found a general agreement between observed galaxies at \textit{z}$\sim$6 and the SED templates used to built \spr{}, confirming the validity of the templates used, even at high-$z$. \par
Looking at other wavelengths, the part of the UV LF dominated by galaxies (i.e. AB magnitude M$_{1600\AA}>$-23) in \spr{} is underestimated at \textit{z}$>$2, indicating that a population of dust-poor galaxy may be missing in the simulation. The absence of such population should however not impact the results presented in this paper focused on future IR missions. \par
The bright-end of the X-ray LF of the simulation is in agreement with previous observational results up to \textit{z}$=$6 \citep{Aird2010,Aird2015,Miyaji2015,Vito2018,Wolf2021}, above which no observations are at the moment available. The faint-end slope, on the contrary, is slightly overestimated and the difference between our simulation and observations increases with redshift, from a factor of four at \textit{z}$\sim$0.5 to a factor of $\sim$15 at \textit{z}$\sim$5, but still within the uncertainties. This will be further investigated in a future work and it is expected to have an impact on the comparison between IR and X-ray fluxes presented in Sec. \ref{sec:xray}. 

\par We can overall conclude that the \spr{} simulation can reliably be used to perform SPICA, GEP and OST predictions. We refer to \citetalias{Bisigello2021} for more information about \spr{}.\par

\begin{table}[]
    \centering
    \caption{The different galaxy populations included in \spr{} and the reference of the LF or GSMF from which their number densities are derived.}
    \resizebox{0.49\textwidth}{!}{
    \begin{tabular}{c|c}
        \hline\hline
        Population & Reference \\
        \hline
        Spiral & \citet{Gruppioni2013}\\ 
        SB & \citet{Gruppioni2013}\\ 
        SF-AGN & \citet{Gruppioni2013}\\ 
        SB-AGN & \citet{Gruppioni2013}\\ 
        AGN1 & \citet{Gruppioni2013},\citetalias{Bisigello2021}\\ 
        AGN2 & \citet{Gruppioni2013},\citetalias{Bisigello2021}\\
        Elliptical & \citet{Cirasuolo2007}, \citet{Arnouts2007}\\
        & and \citet{Beare2019}\\
        Dwarf & \citet{Huertas-Company2016}\\
        \hline\hline
    \end{tabular}}
    \label{tab:LFref}
\end{table}

%%%%%%%%%%%%%%%%%%%%%%%%%%
\subsection{Simulated catalogues}

\begin{figure}
    \centering
    \includegraphics[width=0.89\linewidth,keepaspectratio]{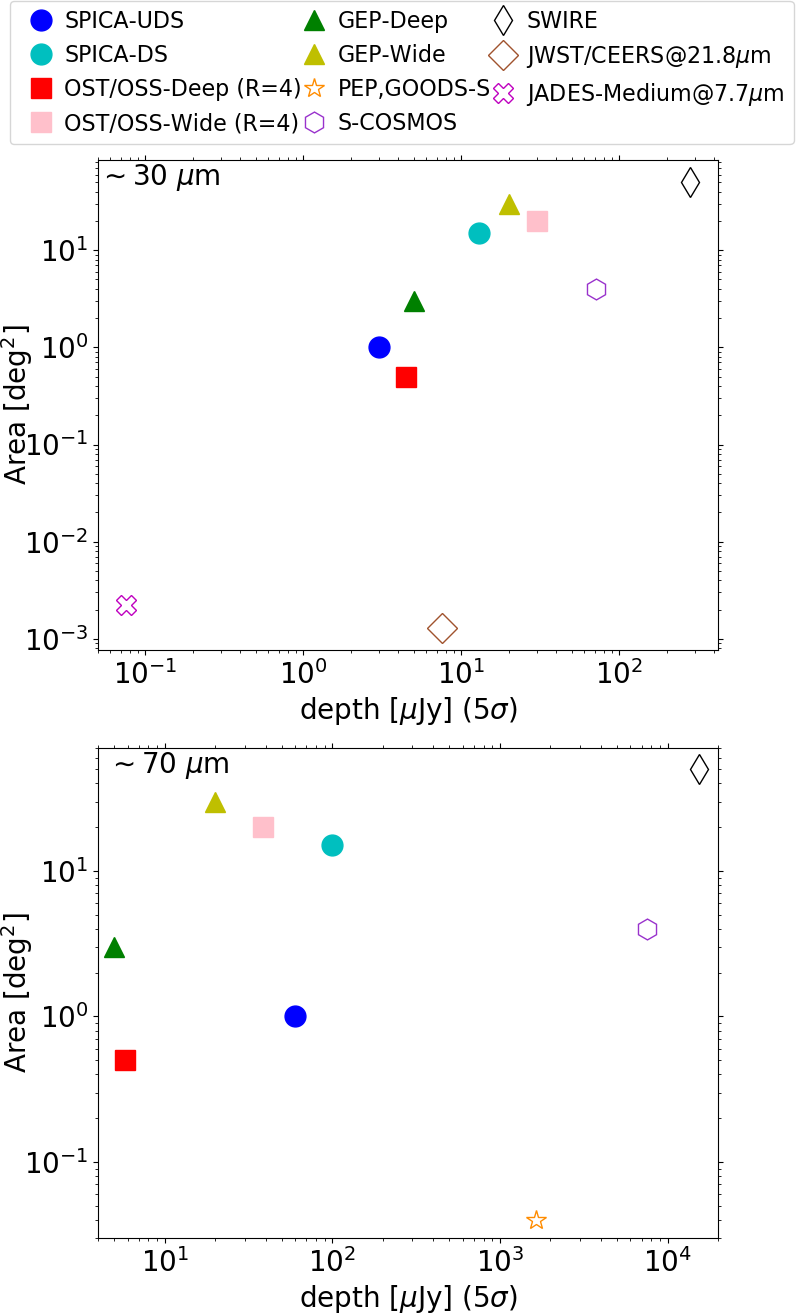}
    \caption{Comparison between area and depth covered by different surveys, observing around 30 (\textit{top}) and 70 $\mu$m (\textit{bottom}). Coloured points indicate the surveys considered in this work (see legend). Empty points indicate some past surveys at 24 and 70 $\mu$m. %: Herschel PEP survey in the GOODS-S at 70 $\mu$m \citep[\textit{orange star};][]{Magnelli2013}, S-COSMOS at 24 and 70 $\mu$m \citep[\textit{purple hexagons};][]{Sanders2007}, SWIRE at 24 and 70 $\mu$m \citep[\textit{black narrow diamonds};][]{Lonsdale2003}. 
    For comparison, we also include two future JWST surveys: the JWST/CEERS survey at 21.8 $\mu$m (\textit{brown diamonds}) and the JWST/JADES medium survey at 7.7 $\mu$m (\textit{magenta crosses}). }
    \label{fig:Area_depth}
\end{figure}

In the next sections we describe in detail the simulated catalogues we created resembling six different surveys of SPICA, OST and GEP. 
To show the improvement of such surveys with respect to previous ones, we show in Figure \ref{fig:Area_depth} the comparison between the depth and area of the considered SPICA, OST and GEP surveys with the depth and area of past surveys observing around 30 and 70 $\mu$m \citep[i.e.][]{Lonsdale2003,Sanders2007,Magnelli2013}. \par
In the same figure, we also include a comparison with two approved James Webb Space Telescope \citep[JWST;][]{Gardner2009} surveys: the Mid Infrared Instrument \citep[MIRI][]{Kendrew2015,Rieke2015,Wright2015} observations at 21.8 $\mu$m which is part of the Cosmic Evolution Early Release Science (CEERS\footnote{\url{https://ceers.github.io}}) survey and the MIRI 7.7 $\mu$m observations of the Medium JWST Advanced Deep Extragalactic Survey (JADES\footnote{\url{https://www.cosmos.esa.int/web/jwst-nirspec-gto/jades}}). JWST and the three far-IR missions analysed in this work are complementary in wavelengths, as JWST covers $\lambda<$28.8 $\mu$m while SPICA, OST and GEP cover $\lambda>$20, 25 and 10 $\mu$m, respectively. At the same time, JWST is particularly suited for performing very deep mid-IR observations, as in imaging it can reach a 10$\sigma$ depth of 7.09 $\mu$Jy at 21.0 $\mu$m in 10$^{4}$ s \citep{Glasse2015}, but they are limited to small area in the sky given the small MIRI field-of-view \citep[74$^{\prime\prime}\times$113$^{\prime\prime}$;][]{Bouchet2015}. 

\subsubsection{SPICA}\label{sec:intro_SPICA}

In this work, we simulated two different spectro-photometric surveys, following two planned SPICA surveys (Table \ref{tab:surveys}):
\begin{itemize}
    \item an Ultra-Deep Survey (UDS) covering 1 deg$^{2}$ in 600 h, reaching down to 3 $\mu$Jy (5$\sigma$) with SMI at 34 $\mu$m and 60 $\mu$Jy (5$\sigma$) with B-BOP at 70 $\mu$m. In the same amount of time SMI-LR would simultaneously obtain spectra down to 39 $\mu$Jy at 20 $\mu$m and 65 $\mu$Jy at 30 $\mu$m (5$\sigma$). 
    \item a deep survey (DS) covering 15 deg$^{2}$ in 600 h, reaching down to 13 $\mu$Jy (5$\sigma$) with SMI at 34 $\mu$m and 100 $\mu$Jy (5$\sigma$) with the B-BOP at 70 $\mu$m. In the same amount of time SMI-LR would simultaneously obtain spectra down to 150 $\mu$Jy at 20 $\mu$m and 250 $\mu$Jy at 30 $\mu$m (5$\sigma$). 
\end{itemize}
Both spectroscopic limits are derived considering a low background level (i.e., 19 MJy/sr at 27 $\mu$m). Photometric confusion noise is not included in \spr{} and corresponds at 5$\sigma$ to 9 $\mu$Jy and 0.25 mJy at 34 $\mu$m and 70 $\mu$m, respectively. The confusion noise may be broken with the help of the available spectroscopic data \citep{Raymond2010}, but it is necessary to consider with caution sources with fluxes similar or below the mentioned confusion noise, as they may need additional de-blending analysis. \par
\begin{figure*}[h!]
    \centering
    \includegraphics[width=0.34\linewidth,keepaspectratio]{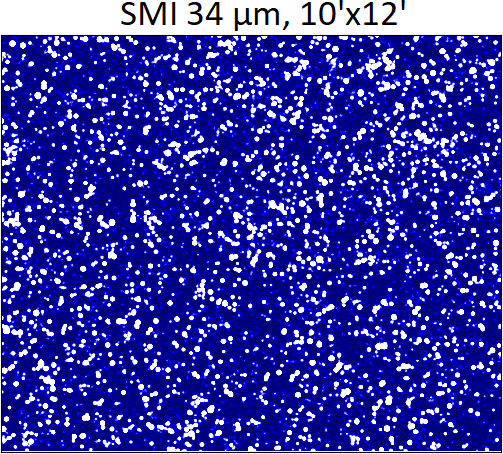}
    \includegraphics[width=0.34\linewidth,keepaspectratio]{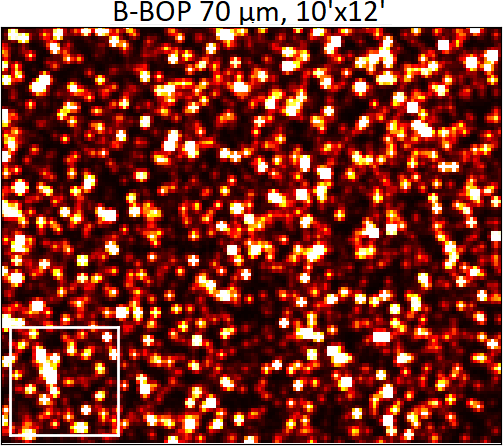}
    \includegraphics[width=0.295\linewidth,keepaspectratio]{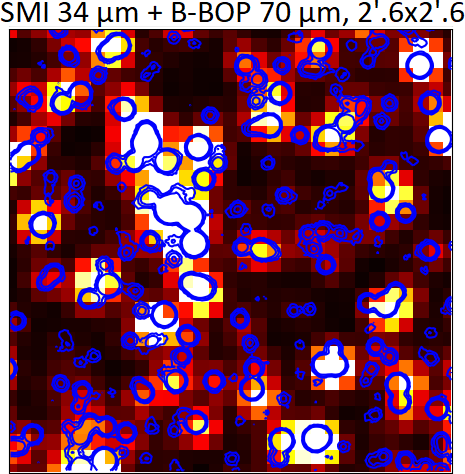}
    \caption{\textit{Left:} Simulated SMI image at 34 $\mu$m of a field of 10$^{\prime}\times$12$^{\prime}$, which corresponds to a single SMI pointing. \textit{Centre:} Simulated B-BOP image at 70 $\mu$m of the same field. The B-BOP field-of-view ($2'.6 \times 2'.6$) is shown for comparison in the bottom left (\textit{white square}). \textit{Right:} Zoom-in of the B-BOP image in the B-BOP field-of-view. Blue contours are the superimposed SMI image (5, 10 and 20$\sigma$). 
    Each point corresponds to one galaxy or blends of several galaxies simulated with \spr{}, depending on source blending, down to the noise level. }
    \label{fig:images}
\end{figure*}

Galaxies in the two mock catalogues are selected to have a signal-to-noise (S/N) larger than 3 either in the SMI or B-BOP photometric filters, similarly to what is done with real data to limit the number of false detection.
In Figure \ref{fig:images} we report as an example a 10$^{\prime}\times$12$^{\prime}$ simulated image with SMI at 34 $\mu$m (\textit{left panel}) and B-BOP at 70 $\mu$m (\textit{central panel}). The chosen area corresponds to a single SMI pointing, whereas a B-BOP pointing covers a smaller area, i.e. 2$^{\prime}$.6$\times$2$^{\prime}$.6. In the right panel we show a zoom-in of the B-BOP pointing with, superimposed, the SMI image. In this figure it is possible to appreciate the different size of the point-spread function (PSF), which has been approximated by a two-dimensional Gaussian. The PSF has a full-width-half-maximum of 3.5$^{\prime\prime}$ at 34 $\mu$m and of 9$^{\prime\prime}$ at 70 $\mu$m. %In Figure \ref{fig:spectra} we instead report some examples of simulated SMI-LR spectra with good S/N($>$50 on average) for different galaxy populations. These examples show the variety of spectra that would be obtained with an instrument build with the specifications of the SMI.

% \begin{figure}[h!]
%     \centering
%     \includegraphics[width=1\linewidth,keepaspectratio]{spectra_example_alltypes.png}
%     \caption{Examples of simulated SMI-LR spectra with average S/N$>$50 for all the considered galaxy populations. We indicate on the right the galaxy type and redshift corresponding to each object. Spectra are arbitrary shifted vertically for clarity. \textcolor{red}{Figura da sistemare}. }
%     \label{fig:spectra}
% \end{figure}

\begin{table}[h!]
    \centering
     \caption{5$\sigma$ depths at four wavelengths corresponding to the two SPICA photometric filters and LR spectrograph at two reference wavelengths, as planned for the two spectro-photometric surveys considered in this work. }
    \begin{tabular}{cccc}
        \hline\hline
        Instrument & $\lambda$ [$\mu$m]& \multicolumn{2}{c}{5$\sigma$ depth [$\mu$Jy]} \\
        & & UDS (1 deg$^{2}$) & DS (15 deg$^{2}$)\\
        \hline
        SMI & 34 &    3 &  13\\
        B-BOP & 70 &    60 &  100 \\
        SMI-LR & 20 & 39 & 150 \\
         & 30 & 65 & 250\\
         \hline\hline
    \end{tabular}
    \label{tab:surveys}
\end{table}

\subsubsection{OST}\label{sec:intro_OST}
The Origins Survey Spectrometer \citep[OSS;][]{Bradford2018} planned for OST consists of six channels covering between 25 and 589 $\mu$m with a resolution of R$=$300. In this work, we compared the two aforementioned SPICA surveys with the Deep and Wide surveys planned for OST (hereafter OST-Deep and OST-Wide), covering, respectively, 0.5 and 20 deg$^{2}$ with a 5$\sigma$ depth of 39.4 and 262.5 $\mu$Jy at 25-44 $\mu$m, at a resolution of R$=$300. \par
For comparison with the two SPICA photometric surveys, we also considered the photometric points derived considering each OST channel as a photometer, i.e. at a resolution of R$=$4 instead of R$=$300. This corresponds to a 5$\sigma$ depth of 4.5 and 30.3 $\mu$Jy at 25-44 $\mu$m in the OST-Deep and OST-Wide survey, respectively. Taking into account the central wavelengths of the different filters, in the part of this work focused on results based on photometric data, we compared results of the OST/OSS Ch1 with the SPICA/SMI filter at 34 $\mu$m and of OST/OSS Ch2 with the SPICA/B-BOP filter 70 $\mu$m. We instead considered all six OST/OSS channels when comparing the spectroscopic capability of the two instruments. \par
Confusion noise should not be a major problem in OST, given the large primary mirror, and it corresponds to 120 nJy and 1.1 mJy at 50 and 250 $\mu$m. As for SPICA, the available spectra can be used to de-blend galaxies and break the confusion noise \citep{Raymond2010}. The full list of OST/OSS channels and their expected observational depths are listed in Table \ref{tab:surveys_OST}.

\begin{table}[h!]
    \centering
     \caption{5$\sigma$ depths at six wavelengths corresponding to the OST/OSS spectroscopic channels, as planned for the two spectroscopic surveys (i.e. OST-Deep and OST-Wide) considered in this work. }
    \begin{tabular}{cccccc}
        \hline\hline
        Channel & $\lambda$ [$\mu$m]& \multicolumn{4}{c}{5$\sigma$ depth [$\mu$Jy]} \\
        & & \multicolumn{2}{c}{Deep (0.5 deg$^{2}$)} & \multicolumn{2}{c}{Wide (20 deg$^{2}$)}\\
        & & R=4 & R=300 & R=4 & R=300\\
        \hline
        Ch1 & 25-44 & 4.5 & 39.4 & 30.3 & 262.5\\
        Ch2 & 42-74 & 5.8 & 50.4 & 38.8 & 336.1\\
        Ch3 & 71-124 & 8.4 & 72.4 & 55.8 & 483.1\\
        Ch4 & 119-208 & 21.8 & 188.9 & 145.5 & 1260\\
        Ch5 & 200-350 & 21.8 & 188.9 & 145.5 & 1260\\
        Ch6 & 336-589 & 70.9 & 614.1 & 472.9 & 4100\\
        \hline\hline
    \end{tabular}
    \label{tab:surveys_OST}
\end{table}

\subsubsection{GEP}\label{sec:intro_GEP}
GEP \citep{Glenn2018} is a NASA concept of a 2 m cold (4 K) telescope with photometric and spectroscopic capability.
The GEP Imager (GEP-I) is planned to have 23 bands covering 10-400 $\mu$m, with spectral resolution R=8 at 10-95 $\mu$m and R=3.5 at 95-400 $\mu$m. Considering the central wavelengths of the different filters, in this work we compared results for the SPICA/SMI filter at 34 $\mu$m and SPICA/B-BOP at 70 $\mu$m with the GEP-I filters centred at $\sim$33 (Band 10) and 73 $\mu$m (Band 16). \par
At the moment of writing, there are four surveys planned: 
\begin{itemize}
    \item An all sky survey with a 5$\sigma$ depth of $\sim$1 mJy.
    \item A survey covering 300 deg$^{2}$ with a 5$\sigma$ depth of $\sim$50 $\mu$Jy.
    \item A survey covering 30 deg$^{2}$ with a 5$\sigma$ depth of $\sim$20 $\mu$Jy (hereafter GEP-Wide).
    \item A survey covering 3 deg$^{2}$ with a 5$\sigma$ depth of $\sim$5 $\mu$Jy (hereafter GEP-Deep).
\end{itemize}
For a better comparison with the SPICA surveys, in this work we considered only the last two surveys. Given the size of the primary mirror, GEP is expected to reach the confusion noise level in imaging at approximately 70 $\mu$m in the two considered surveys \citep{Glenn2019}. \par
In addition, the GEP Spectrometer (GEP-S) concept consists of four gratings covering from 24 to 193 $\mu$m with R=200. However, current plans for GEP-S include only pointed observations \citep{Glenn2019}. Therefore, we decided to limit the comparison of GEP and SPICA only to their photometric capability.\par

%%%%%%%%%%%%%%%%%%%%%%%%%%%%%%%%%%%%%%%%%%%%%%%%%%%%%%%%%%%%%%%%%%%%%%%%%%%%%%%%%%%%%%%%%%%%%%%%%

\section{Predictions}\label{sec:pred}
\subsection{Galaxy populations expected from photometric surveys} % not good, improve the title
In this section, we give an overview of the physical properties of the galaxy populations that are expected to be detected by the considered photometric surveys, as derived from \spr{}. We concentrate on the nature of such objects, but we highlight that, for their analysis, it may be necessary to rely on additional spectroscopic data (see Sec. \ref{sec:spectra}) or a larger set of photometric data including also other wavelengths.\par

%%%%%%%%%%%%%%%%%%%%%%%
\subsubsection{IR luminosities}

\begin{figure*}[h!]
\begin{center}
\includegraphics[width=1\linewidth,keepaspectratio]{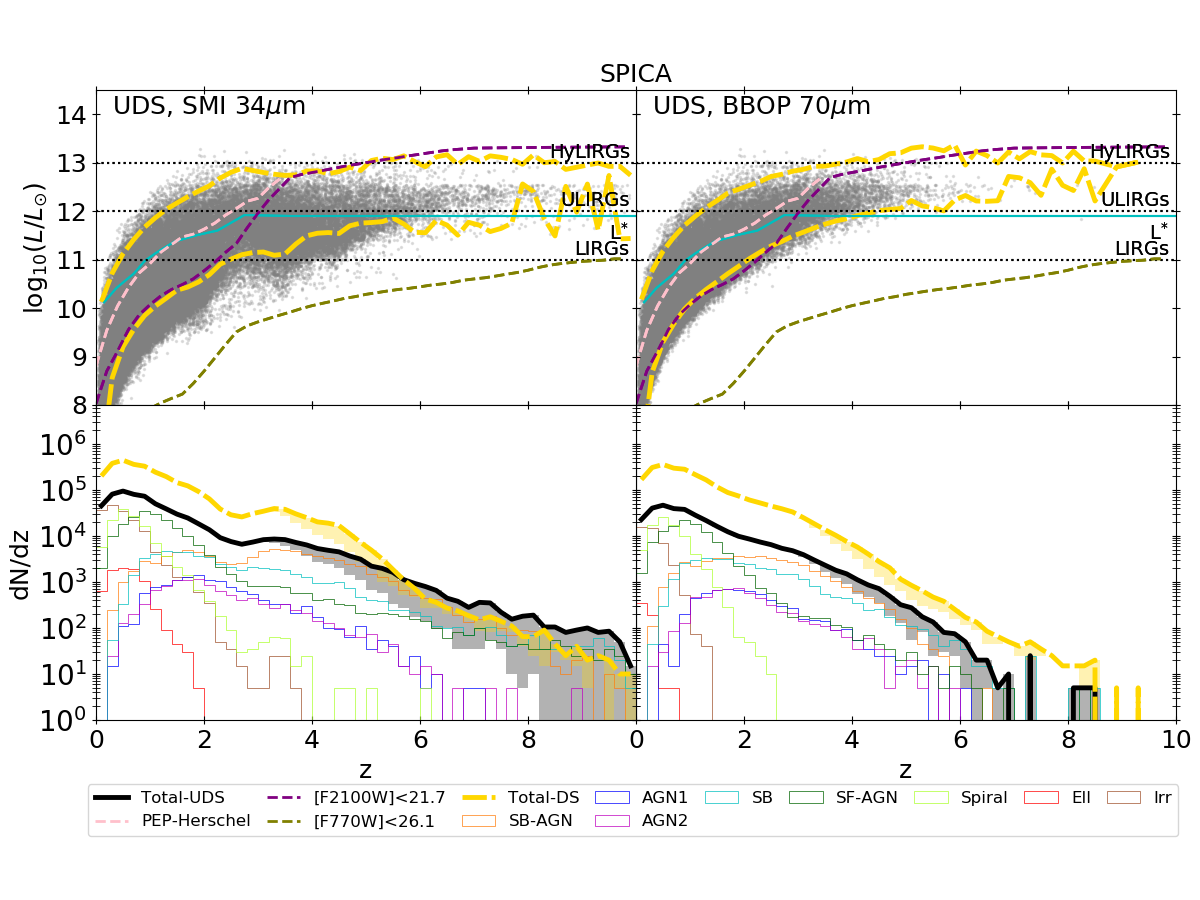}
\caption{\textit{Top:} L$_{IR}$ vs. redshift for simulated galaxies (\textit{grey points}) detected at 34 $\mu$m (\textit{left}) and 70 $\mu$m (\textit{right}) in the SPICA UDS.  %The various galaxy populations are shown in different colours: spirals (\textit{light-green}), SB (\textit{cyan}), SF-AGN (\textit{dark-green}), SB-AGN (\textit{orange}), AGN1 (\textit{blue}), AGN2 (\textit{magenta}), elliptical  (\textit{red}) and irregulars (\textit{brown}). 
The \textit{dashed tick yellow lines} refer to the L$_{IR}$ area occupied by galaxies detected in SPICA DS. In each panel, the cyan solid line indicates the luminosity of the knee of the LF, as derived by \citet{Gruppioni2013}, while the \textit{pink dotted line} show the 80$\%$ completeness of the $Herschel$-PEP survey at 100 $\mu$m in the GOODS-S \citep{Berta2013}. We also report the 80$\%$ completeness of the two samples with JWST/MIRI [F2100]<21.7 and [F770W]<26.7. The horizontal dotted black lines indicate the IR luminosity limit of HyLIRGs, ULIGRs and LIRGs. Results are obtained considering the high-$z$ extrapolation with $k_{\Phi}=$-1. Both surveys would have allowed to observe not only ULIRGs, as done with \hers, but also less luminous galaxies.
\textit{Bottom:} Redshift distribution per unit redshift interval of sources detected at 34 $\mu$m with SMI (\textit{left}) and at 70 $\mu$m with B-BOP, as expected in the UDS (\textit{tick black lines}) and in the DS (\textit{tick yellow dashed lines}).  For the UDS, we report also the distributions of the different sub-populations (\textit{coloured lines}), all derived considering $k_{\Phi}=$-1. Results obtained with other \textit{z}$>$3 extrapolations, i.e. from $k_{\Phi}=$-4 to -2, are included in the grey and yellow areas. Overall, observations at 34 $\mu$m could reach galaxies up to \textit{z}$\sim$8.} \label{fig:zdist_SPICA}
\end{center}
\end{figure*}

\begin{figure*}[h!]
\begin{center}
\includegraphics[width=1\linewidth,keepaspectratio]{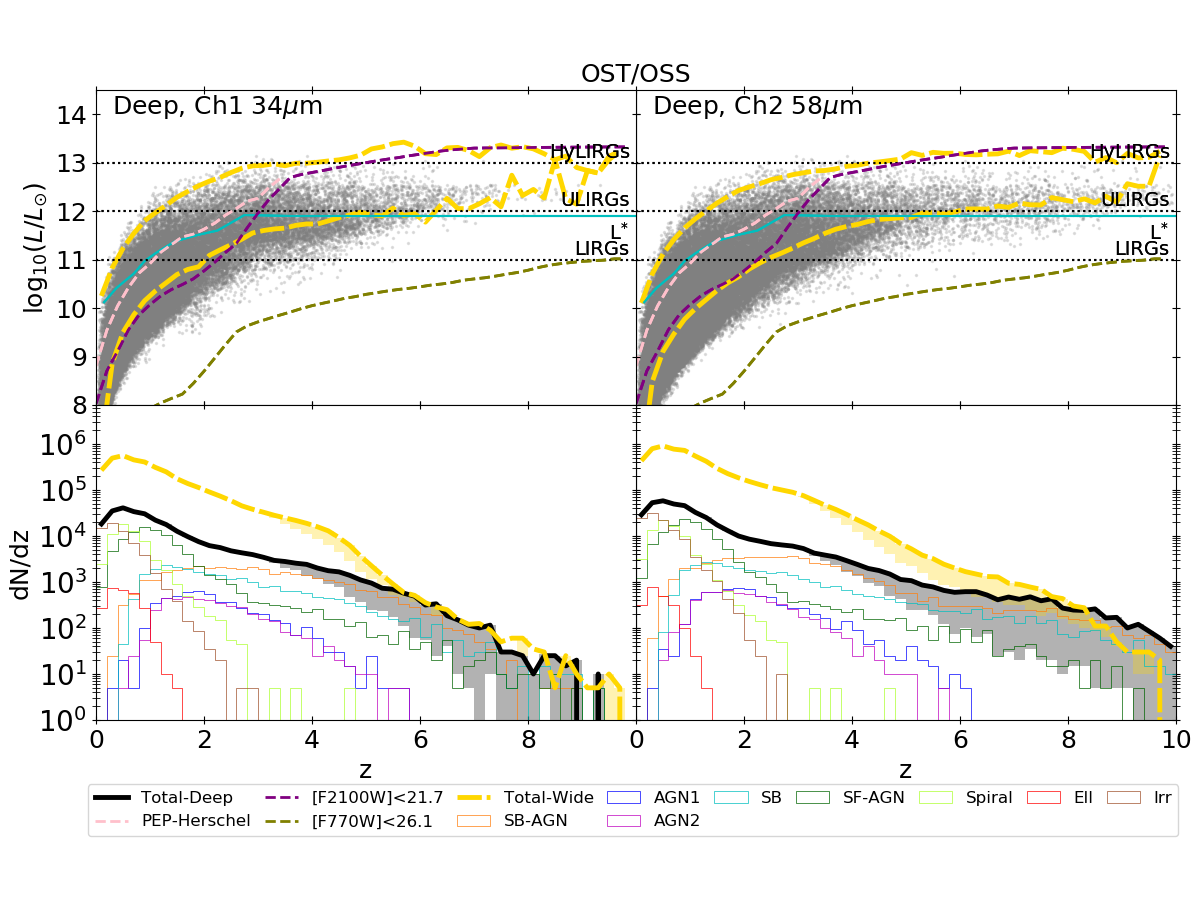}
\caption{Same as Figure \ref{fig:zdist_SPICA}, but for galaxies detected with OST in channel 1 and 2 in the OST-Deep survey. \textit{Tick yellow dashed lines} show the results for the OST-Wide survey. Both surveys will allow the detection of  galaxies  up to \textit{z}$=$7, thus extending the existing observations well below the ULIRG typical luminosity, at least up to \textit{z}$=$5.}\label{fig:zdist_OST}
\end{center}
\end{figure*}

\begin{figure*}[h!]
\begin{center}
\includegraphics[width=1\linewidth,keepaspectratio]{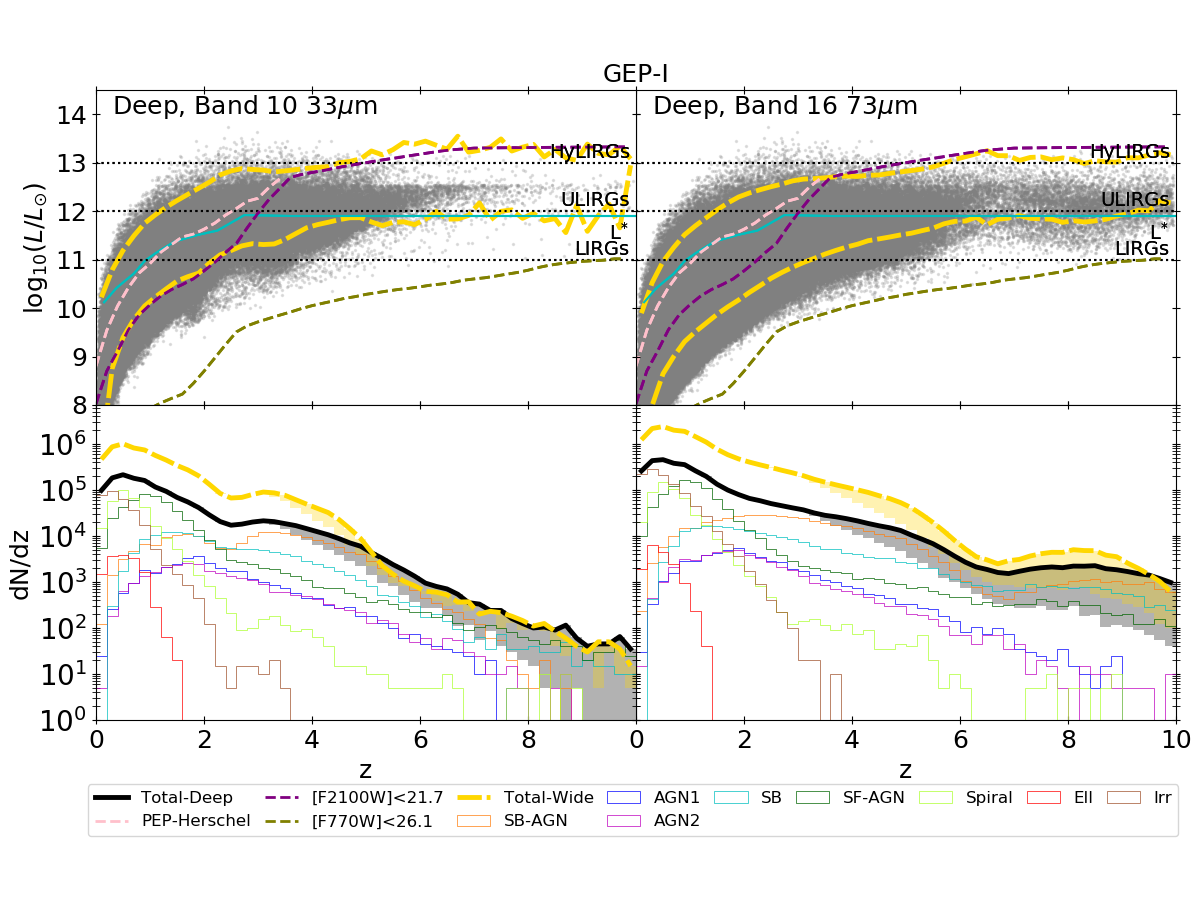}
\caption{Same as Figure \ref{fig:zdist_SPICA}, but for galaxies detected with GEP-I in channel 10 and 16 in the GEP-Deep survey. \textit{Tick yellow dashed lines} show the results for the GEP-Wide survey. At both wavelengths, these surveys will allow the detection of  galaxies up to \textit{z}$=$9  and,  at 73 $\mu$m, of some galaxies below the ULIRG luminosity at \textit{z}$>$5. }\label{fig:zdist_GEP}
\end{center}
\end{figure*}

We start by investigating the IR luminosity of the galaxies that we expect to observe with a SPICA-like mission (Figure \ref{fig:zdist_SPICA}), compared to predictions for OST (Figure \ref{fig:zdist_OST}) and GEP (Figure \ref{fig:zdist_GEP}). \par
In both the UDS and DS, SMI would be able to detect galaxies with IR luminosities at the knee of the luminosity function, as derived by \citet{Gruppioni2013}, up to \textit{z}$=$6. This would be possible, at least for some objects, even up to \textit{z}$=$10, while it would be limited to \textit{z}$<$4 for B-BOP observations in both surveys.\par
The depth and area of the DS are ideal to statistically investigate the population of Ultra-luminous IR (ULIRG) and Hyper-luminous IR galaxies (HyLIRG), i.e. respectively $L_{IR}>$10$^{12}\,L_{\odot}$ and $L_{IR}>$10$^{13}\,L_{\odot}$. We indeed would expect to detect around 4-6$\times$10$^{4}$ ULIRGs up to redshift \textit{z}$=$10, showing a mean redshift in the range \textit{z}$=$2.8-3.5 and $\sim$700-900 HyLIRGs with a mean redshift \textit{z}$=$2.6-3.8. For comparison, in the UDS, which covers only 1 deg$^{2}$, we would expect to detect only a few tens (30-50) of HyLIRGs. For an area of 100 deg$^{2}$ and considering the same observational depth of the DS, which was not possible with SPICA but may be feasible with future IR missions, we would expect $\sim$5000-6000 HyLIRGs with mean redshift \textit{z}$=$3.6-3.8 and 3.5-4.7$\times$10$^{5}$ ULIRGs with mean redshift \textit{z}$=$3.0-3.4. \par
The uncertainties on the numbers of ULIRG and HyLIRG are due to the extrapolation of the IR LF at \textit{z}$>$3, which impacts also the mean redshift of the simulated sample. In fact, if we consider $k_{\Phi}=-$4 instead of $k_{\Phi}=-$1, the average redshift decreases as there are fewer galaxies at \textit{z}$>$3. \par
The predicted HyLIRGs number (i.e. 30-50 deg$^{-2}$) corresponds to a larger surface density than the one observed by \citet{Wang2021}, which ranges between 5 and 18 HyLIRGs per deg$^{2}$. However, these surface densities are consistent with each other, once we take into account the uncertainties on the bright-end of the IR \hers LF by \citet{Gruppioni2013}, which varies between 0.07 and 0.43 dex for $L_{IR}>$10$^{13}\,L_{\odot}$, depending on the considered redshift. Even considering the more optimistic estimation of the HyLIRGs surface density, we need an area larger than $\sim$70 arcmin$^{2}$ to detect at least one of them. This consideration shows the complementarity of missions such as SPICA, OST or GEP with respect to JWST, for which it is difficult to observe large areas, particularly in the mid-IR. \par
The galaxy population observed by a SPICA-like mission would not be limited to extreme IR galaxies. Indeed, with SMI at 34 $\mu$m, we would expect to observe below the luminosity of the so-called Luminous IR galaxies (LIRGs, L$_{IR}=10^{11}\,{\rm L}_{\odot}$) up to \textit{z}$=$4.0 and \textit{z}$=$2.5 in the UDS and DS, respectively. Even if these observations would be mainly limited to \textit{z}$<$2.0 at 70 $\mu$m, they would lead to breakthrough results such as the unprecedented, direct measure of the faint-end slope of the luminosity function at these wavelengths. \par
%%%%
The OST/OSS Ch1 depth in the OST-Deep survey is similar to the SMI depth at 34 $\mu$m in the SPICA UDS and, indeed, the IR luminosity of observed galaxies is expected to be similar. In the OST-Wide Survey the OST/OSS Ch1 is instead planned to be shallower than SMI at 34 $\mu$m, i.e. 30.3 $\mu$Jy compared to 13 $\mu$Jy. This limits the number of the detectable faint galaxies with respect to what was expected for SPICA. However, the planned observational depths for the OST/OSS Ch2 are deeper than the ones planned for SPICA/B-BOP at 70 $\mu$m. For this reason, in the OST/OSS Ch2 we expect to observe galaxies below the ULIRG limit up to \textit{z}$\sim$7.5 in the OST-Deep Survey and up to \textit{z}$\sim$5.5 in the OST-Wide Survey. \par
%%%%
GEP-I observational depths are expected to be of $\sim$5 and 20 $\mu$Jy, in the 3 and 30 deg$^{2}$ surveys, respectively. Such surveys are therefore deeper than both SPICA surveys at 70 $\mu$m and the SPICA DS at 34 $\mu$m. For this reason, we expect to see galaxies with IR luminosity below the ULIRGs regime up to \textit{z}$=$5.5 (5.0) and 8.5 (4.0) at 33 and 73 $\mu$m in the deepest (widest) considered GEP survey. However, we note that the number and the maximum observable redshift of galaxies with L$_{IR}<10^{12}\,{\rm L}_{\odot}$ detected by GEP-I may be underestimated, as we consider only two of the 23 filters available. A similar case does not hold for OST, as the most sensitive filters are those at the shortest wavelengths and they are the filters we considered in this work.\par

We also compared the 80$\%$ limits of the L$_{IR}$, derived using \spr{}, of two samples selected in two JWST/MIRI filters: [F2100W]<21.7 (i.e. 7.5 $\mu$Jy) and [F770W]<26.7 (i.e. 0.07 $\mu$Jy). The two selections correspond to the magnitude limits of the CEERS and JADES surveys, respectively, in their two longest wavelength filters (see Sec. \ref{sec:intro_SPICA}). For this comparison, photometric errors are not included in the JWST fluxes. \par
The first flux cut corresponds to a L$_{IR}$ limit similar to the UDS and DS limits at 70 $\mu$m, to the OST-Wide limit at 34 $\mu$m and to the GEP-Wide limit at 33 $\mu$m, at least up to \textit{z}$=$3, while it probes galaxies with brighter L$_{IR}$ than the other SPICA, OST and GEP surveys.  The second JWST flux threshold corresponds to a L$_{IR}$ limit much fainter than all the considered SPICA, OST and GEP surveys, showing the power of JWST to observe very faint galaxies. \par
The expected number of galaxies observed by JWST will be smaller than the number observed by the considered SPICA, OST and GEP surveys, given the small sky area covered by the two considered JWST surveys (i.e. $\sim$4.6 and 8 arcmin$^{2}$ for the CEERS F2100W and for the JADES F770W observations, respectively). These small areas would also limit or preclude observations of very bright galaxies. In addition, JWST does not allow one to fully sample the  far-IR dust bump at any redshift, but this is instead possible with OST and GEP, enabling the characterisation of the properties of the dust in high-redshift galaxies. Moreover, the same would have been possible with SPICA, thanks to the capabilities of 
the B-BOP instrument.

%%%%%%%%%%%%%%%%%%%%%%%
\subsubsection{Redshift distribution}

In Figure \ref{fig:zdist_SPICA} we show the expected redshift distribution per unit redshift interval of the galaxies that would be detected in the SPICA UDS and DS with SMI at 34 $\mu$m and B-BOP at 70 $\mu$m. In Table \ref{tab:zcounts} we list the predicted number of galaxies that would be detected in the considered filters in the UDS and DS in different redshift intervals. At \textit{z}$<$2, we would expect a few times 10$^{4}$ objects per redshift bin to be detected at both wavelengths in the UDS, while this number increases to a few times 10$^{5}$ in the DS. These are predominantly normal star-forming galaxies, i.e. labelled as spiral, irregular or SF-AGN in the figures, with an increment of the starburst fraction with increasing redshift. Such observations would have allowed for outstanding improvements of our knowledge of the LF, not only at its knee, thanks to the large number of observed galaxies, but also at the faint end, considering the possible observations of galaxies below the LIRG limit. \par
Up to \textit{z}$=$2, we would also expect to observe a sample of elliptical galaxies, whose analysis in the IR has been limited up to now mainly to nearby objects \citep[e.g.][]{Smith2012}. Their numbers would vary between 10$^{2}$ and 10$^{3}$ objects in total, depending on the instrument considered and the area covered. \par
At increasing redshift, the number of observed galaxies would decrease, but we would expect to detect galaxies up to extremely high-$z$ at 34 $\mu$m and at least up to \textit{z}$=$7-8 at 70 $\mu$m, thanks to the combination of depth and area coverage. At 34 $\mu$m, we would expect $\sim$100-900 objects at \textit{z}$>$6 detected in the UDS and slightly less, i.e. $\sim$300-500, in the shallower DS. The uncertainties are due to the high-$z$ extrapolation. The 70 $\mu$m filter would have been less sensitive than the 34 $\mu$m filter, resulting on fewer observed objects. In particular, we predict at maximum a few tens of detections in the UDS and around one hundred in the DS. \par
Keeping in mind the limitations of the simulation at \textit{z}$>$6, we would expect these high-$z$ galaxies to be SBs or composite galaxies with an AGN component (i.e., SB-AGN and SF-AGN), with some AGN-dominated system only in the DS. In the UDS (DS) these galaxies would have an average stellar mass of log$_{10}(\rm{M^*}/\rm{M}_{\odot})=$10.7 (11.1), with values ranging from 8.5 (9.1) to 11.8 (12.2). The average SFR, derived combining UV and IR estimations, would be log$_{10}(\rm{SFR_{UV+IR}}/\rm{M}_{\odot}\,\rm{yr}^{-1})=$2.1 in the UDS and 2.3 in the DS, but it could even reach up to log$_{10}(\rm{SFR_{UV+IR}}/\rm{M}_{\odot}\,\rm{yr}^{-1})=$3.8. However, these SFRs need to be considered as lower limits, as comparing the simulation with observations at \textit{z}$>$6, we derived that a population of highly star-forming systems (i.e. log$_{10}$(sSFR/yr)$\gtrsim$-7.5) may be missing \citepalias[see ][]{Bisigello2021}. The gas-phase metallicity, derived considering the mass-metallicity relation by \cite{Wuyts2014}, would be on average 12+log$_{10}$(O$/$H)$=$8.3-8.4, depending on the considered survey, but it could go as down as 7.5. All these values are derived considering the high-$z$ extrapolation with $k_{\Phi}=$-1. \par
The improvement with respect to previous surveys is outstanding, given that \hers, for example, has been mainly limited to galaxies below $z\leq$3.5 \citep[e.g.][]{Gruppioni2013}. This shows the need for similar surveys when investigating the evolution of both star-formation and black hole accretion in galaxies over a large redshift range. \par
%%%%%%%%%
In Figure \ref{fig:zdist_OST} we show the results for OST/OSS. On one hand, with the OST/OSS Ch1 filter, we expect to be limited to \textit{z}$<$8 in both the OST-Deep and OST-Wide surveys, with the possibility to observe few tens (300) of galaxies \textit{z}$>$8 (\textit{z}$>$6) in the OST-Deep survey if we consider the flattest redshift evolution of the LF (i.e. k$_{\Phi}=-1$). On the other hand, with the OST/OSS Ch2 filter we expect to observe 0.8-2.1$\times$10$^{3}$ galaxies with \textit{z}$>$6 in the OST-Wide survey of 20 deg$^{2}$, depending on the considered extrapolation of the LF at high redshift.\par
In the OST-Deep (OST-Wide) survey such high-$z$ (\textit{z}$>$6) galaxies are expected to have, on average, log$_{10}(\rm{M*}/\rm{M}_{\odot})=$10.5 (11.0) and log$_{10}(\rm{SFR_{UV+IR}}/\rm{M}_{\odot}\,\rm{yr}^{-1})=$2.0 (2.4), considering the $k_{\Phi}=$-1 high-$z$ extrapolation. These galaxies are, on average, less massive and more star-forming than the ones that would be detected by SPICA, because OST/OSS Ch2 observations are planned to be deeper than B-BOP observations. We remind the reader that for a fair comparison with SPICA capabilities we have considered only the first two OST/OSS channels. However, because of the positive k-correction (i.e. rest-frame wavelengths moving closer to the peak of dust emission at increasing redshift), the number of \textit{z}$>$6 galaxies detected by OST is expected to increase when considering all six channels.\par

%%%%%%%%%
Figure \ref{fig:zdist_GEP} shows the results for GEP. In the survey of 3 deg$^{2}$, we instead expect to observe 1-7$\times$10$^{3}$ galaxies at \textit{z}$>$6 with the narrow filter centred at 73 $\mu$m (Band 16) and 200-3500 with the filter centred at 33 $\mu$m (Band 10). In the shallower survey of 30 deg$^{2}$ we instead expect $\sim$500-800 galaxies detected in Band 10 and between 2700 and 1.3$\times$10$^{4}$ in Band 16. \par
In the GEP-Deep (GEP-Wide) survey such galaxies are expected to have, on average, log$_{10}(\rm{M*}/\rm{M}_{\odot})=$10.5 (10.9) and log$_{10}(\rm{SFR_{UV+IR}}/\rm{M}_{\odot}\,\rm{yr}^{-1})=$1.9 (2.2), considering the $k_{\Phi}=$-1 high-$z$ extrapolation. In the GEP-Deep survey we also expect some AGN-dominated galaxies (i.e. AGN1 and AGN2) too rare for the OST-Deep survey or the UDS.  \par
Differences between SPICA, OST and GEP are driven by the different observational depths and different survey sizes, but also, in particular in the case of GEP, by the different filter widths.

\begin{table*}[h!]
    \centering
    \begin{tabular}{c|cc|cc}
        \hline\hline
         Survey & \multicolumn{2}{c|}{UDS}& \multicolumn{2}{c}{DS} \\
         Filter & SMI & B-BOP & SMI & B-BOP\\ 
         \hline
         0$\leq$\textit{z}$<$1 & 7.5$\times$10$^{4}$ & 3.7$\times$10$^{4}$ & 3.4$\times$10$^{5}$ & 2.8$\times$10$^{5}$ \\
         1$\leq$\textit{z}$<$2 & 3.2$\times$10$^{4}$& 1.8$\times$10$^{4}$ & 1.6$\times$10$^{5}$ & 1.3$\times$10$^{5}$ \\
         2$\leq$\textit{z}$<$3  & 8.8$\times$10$^{3}$ & 6.5$\times$10$^{3}$ & 3.7$\times$10$^{4}$ & 4.6$\times$10$^{4}$ \\  
         3$\leq$\textit{z}$<$4 & 7.7$\times$10$^{3}$ (5.8$\times$ 10$^{3}$) & 2.5$\times$10$^{3}$ (1.9$\times$10$^{3}$) & 3.3$\times$10$^{4}$ (2.5$\times$10$^{4}$) & 1.5$\times$10$^{4}$ (1.2$\times$10$^{4}$) \\
         4$\leq$\textit{z}$<$5 & 4.3$\times$10$^{3}$ (1.9$\times$ 10$^{3}$) & 700 (340) & 1.5$\times$10$^{4}$ (6.6$\times$10$^{3}$) & 3.1$\times$10$^{3}$ (1.6$\times$10$^{3}$) \\
         5$\leq$\textit{z}$<$6 & 1.6$\times$10$^{3}$ (400) &  150 (40) & 2.4$\times$10$^{3}$ (950) & 520 (270) \\
         6$\leq$\textit{z}$<$8 & 760 (110) & 30 (2) & 420 (290) & 130 (90) \\
         8$\leq$\textit{z}$<$10 & 180 (2) & 3 (0) & 70 (30) & 12 (10) \\
         \hline\hline
    \end{tabular}
    \caption{Approximated number of galaxies in different redshift intervals detected by SMI and B-BOP, as expected for the UDS and DS. Numbers in brackets are for the most conservative extrapolation at \textit{z}$>$3 considered in this work (i.e. $k_{\Phi}=$-4). }
    \label{tab:zcounts}
\end{table*}

%%%%%%%%%%%%%%%%%%%%%%%
\subsubsection{X-ray fluxes}\label{sec:xray}

\begin{figure}[h!]
    \centering
    \includegraphics[width=1\linewidth,keepaspectratio]{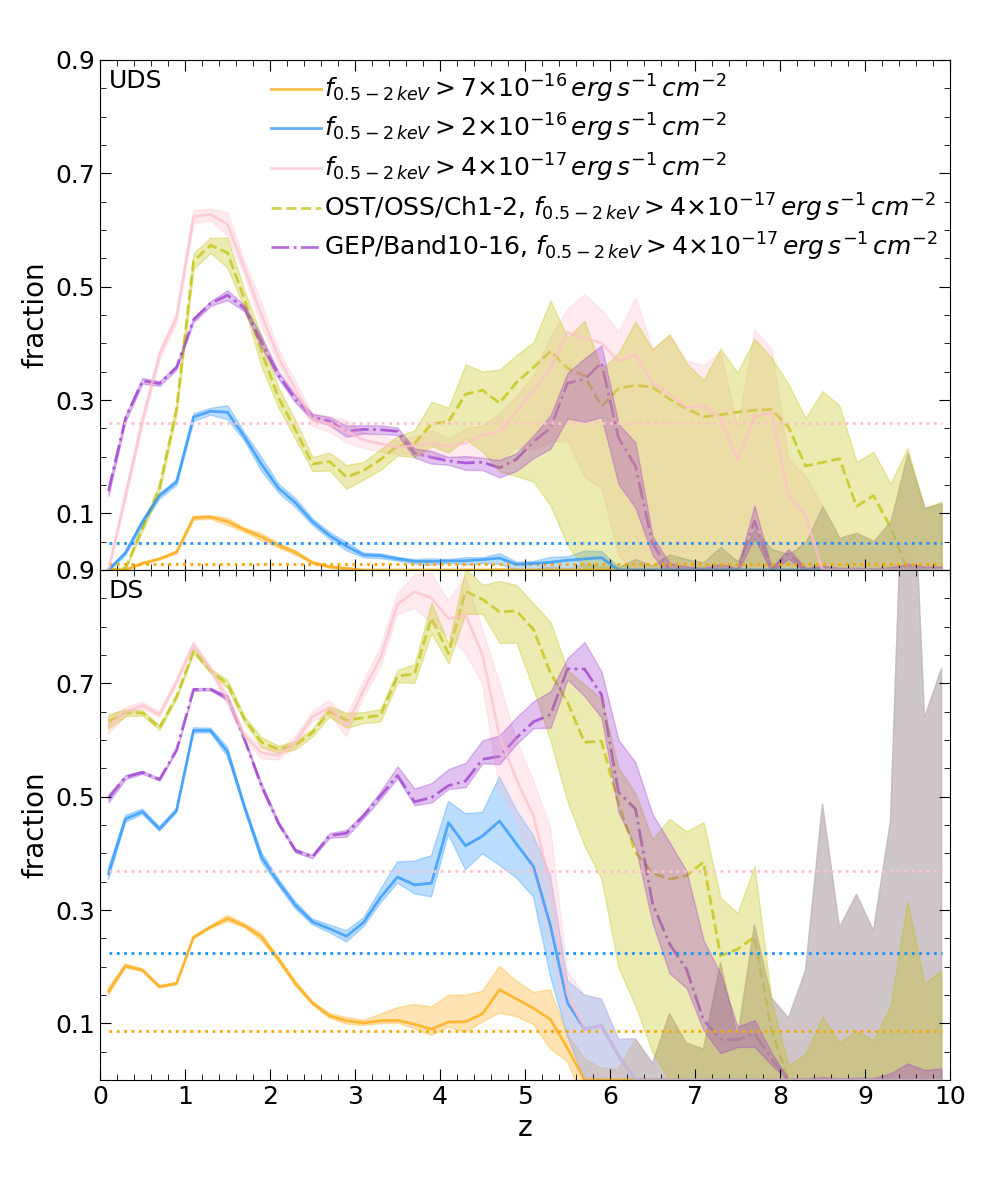}
    \caption{Fraction of simulated AGN detected in the IR with X-ray flux at 0.5-2 keV above 4$\times10^{-17}$ (\textit{pink lines}), 2$\times10^{-16}$ (\textit{blue lines}) and 7 $\times10^{-16}\,{\rm erg\,s}^{-1}\,{\rm cm}^{-2}$ (\textit{orange lines}), in the SPICA UDS (\textit{top}) and DS (\textit{bottom}). We also report the fraction of simulated AGN with X-ray flux at 0.5-2 keV above 4$\times10^{-17}\,{\rm erg\,s}^{-1}\,{\rm cm}^{-2}$ detected with OST/OSS (\textit{green dashed line}) in the OST-Deep (\textit{top}) and OST-Wide survey (\textit{bottom}). The same fraction is shown for AGN detected with GEP (\textit{purple dot-dashed line}) in the survey of 3 (\textit{top}) and 30 deg$^{2}$ (\textit{bottom}). Shaded areas show the uncertainties, including Poisson errors and the extrapolation at \textit{z}$>$3, while horizontal dotted lines indicate the average fraction for SPICA across the entire redshift range. IR observations are key to putting together a complete picture of AGN activity, as X-ray observations may miss a large fraction of them because of dust absorption.}
    \label{fig:Xdet_soft}
\end{figure}

\begin{figure}[h!]
    \centering
    \includegraphics[width=1\linewidth,keepaspectratio]{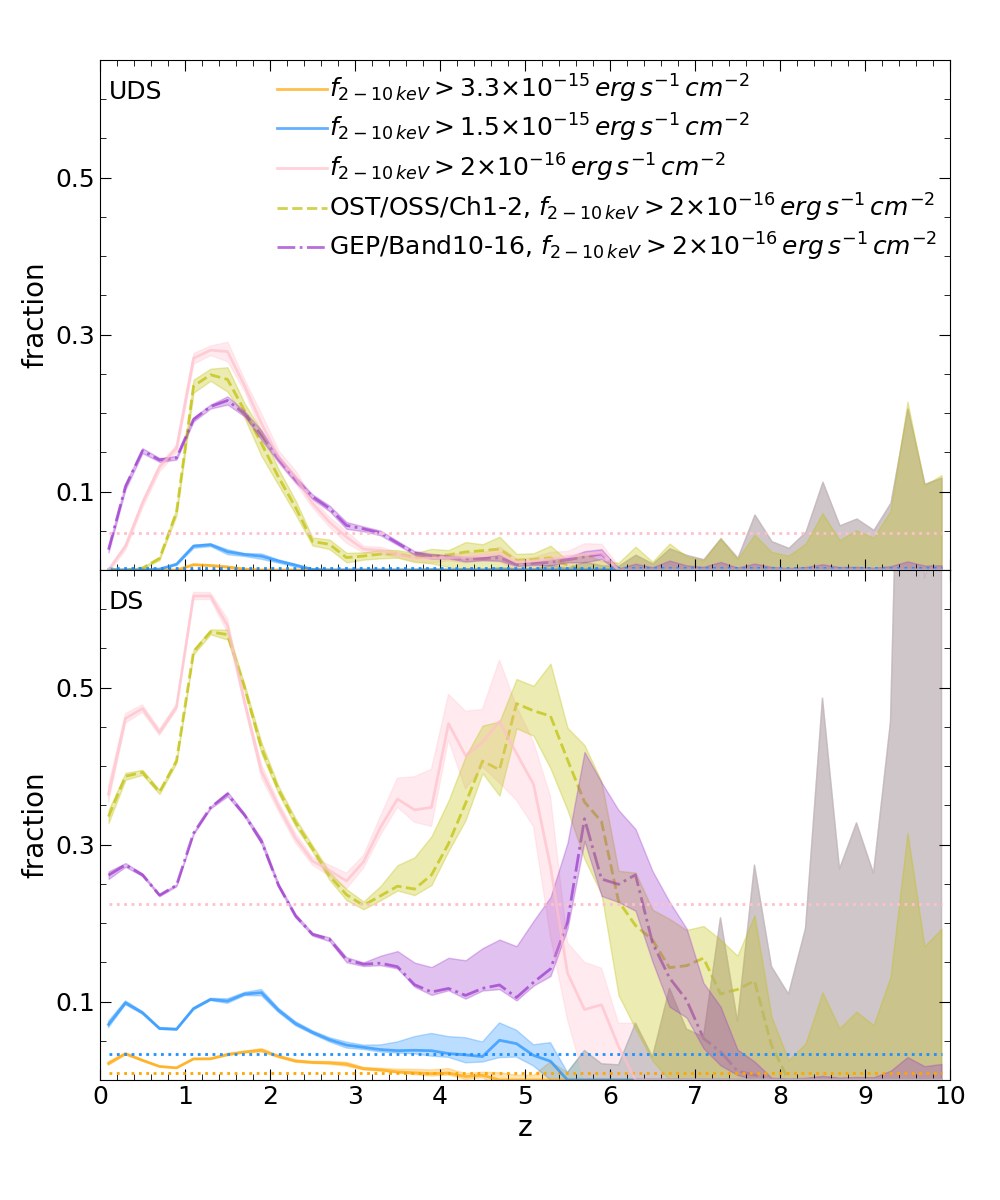}
    \caption{Same as Figure \ref{fig:Xdet_soft}, but for AGN with X-ray flux at 2-10 keV above 3.3$\times10^{-15}$ (\textit{pink lines}), 2.5$\times10^{-15}$ (\textit{blue lines}) and 2 $\times10^{-16}\,{\rm erg\,s}^{-1}\,{\rm cm}^{-2}$ (\textit{orange lines}).}
    \label{fig:Xdet_hard}
\end{figure}

\begin{figure}[h!]
    \centering
    \includegraphics[width=1\linewidth,keepaspectratio]{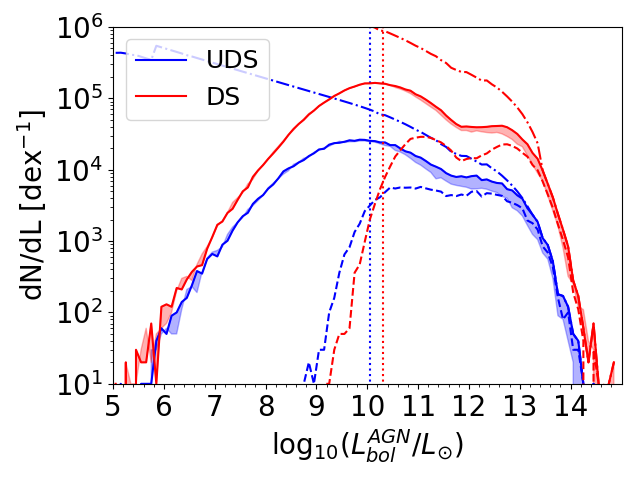}
    \caption{Distribution of the bolometric luminosity of the AGN detected in the SPICA UDS (\textit{blue solid line}) and DS (\textit{red solid line}). The dashed histograms indicate the distribution of the AGN at \textit{z}$>$3. Shaded areas show the uncertainties due to the \textit{z}$>$3 extrapolation. Vertical dotted lines are the median bolometric luminosities of the two surveys across the entire redshift range. The dashed-dotted lines indicate the intrinsic distribution of the bolometric luminosity, before applying any flux selection but normalised to the observed area.} % write better the unit of the y-axis.
    \label{fig:Lbol}
\end{figure}
If we focus on the AGN activity, we can investigate what are the expected soft (0.5-2 keV, Figure \ref{fig:Xdet_soft}) and hard (2-10 keV, Figure \ref{fig:Xdet_hard}) X-ray fluxes for each galaxy detected by a SPICA-like mission, OST or GEP. This is important because, for galaxies with both IR and X-ray observations, it will be possible to analyse the global energetics of the AGN. In addition, IR observations are complementary to X-ray ones \citep{Barchiesi2021}, as they allow observations of the most dust-obscured AGN
that are difficult to detect in the X-rays. As mentioned in Sec. \ref{sec:spritz}, the observed faint end of the X-ray LF is overestimated by the predictions obtained with \spr{}. Therefore, the fractions of IR-detected AGN above certain X-ray fluxes have to be considered as upper limits, at least for the deepest X-ray threshold considered. \par
We compared the expected soft (hard) X-ray fluxes of IR detected AGN with the depth of current and future X-ray surveys: 7$\times10^{-16}\,{\rm erg\,s}^{-1}\,{\rm cm}^{-2}$ (3.3$\times10^{-15}\,{\rm erg\,s}^{-1}\,{\rm cm}^{-2}$) like the XMM-Newton Deep survey in COSMOS \citep{Hasinger2007}, 2$\times10^{-16}\,{\rm erg\,s}^{-1}\,{\rm cm}^{-2}$ (1.5$\times10^{-15}\,{\rm erg\,s}^{-1}\,{\rm cm}^{-2}$) like the Chandra Deep survey in COSMOS-Legacy \citep{Civano2016} and 4$\times10^{-17}\,{\rm erg\,s}^{-1}\,{\rm cm}^{-2}$ (2.0$\times10^{-16}\,{\rm erg\,s}^{-1}\,{\rm cm}^{-2}$) like the planned  \textsl{Athena} \citep{Nandra2013} Deep survey. \par
In the SPICA UDS, we expect, on average, that 26$\%$, 5$\%$ and 1$\%$ of the detected AGN would have soft X-ray fluxes above 4$\times10^{-17}$, 2 $\times10^{-16}$ and 7$\times10^{-16}\,{\rm erg\,s}^{-1}\,{\rm cm}^{-2}$, respectively. This fraction generally would have a peak around \textit{z}$\sim$1.3 and then it would decrease with increasing redshift, going from $\sim$65$\%$ to less than 30$\%$ of galaxies having f$_{0.5-2keV}>4 \times10^{-17}\,{\rm erg\,s}^{-1}\,{\rm cm}^{-2}$ at \textit{z}$>$2.5. \par
In the DS, the fractions of detected AGN with soft X-ray fluxes above the considered limits would be larger than in the UDS, i.e. 37$\%$, 22$\%$ and 9$\%$, on average, of detected AGN with soft X-ray fluxes above 4$\times10^{-17}$, 2$\times10^{-16}$ and 7 $\times10^{-16}\,{\rm erg\,s}^{-1}\,{\rm cm}^{-2}$, respectively. The fraction of galaxies above each X-ray flux cut would be large ($\sim$15, 25 and 70$\%$) even at low redshift, reflecting the reduced number of faint AGN in the DS survey. 
At \textit{z}$>$4-5, the fraction decreases with increasing redshift reaching fractions below 30$\%$, considering the lowest flux cut f$_{0.5-2keV}>4 \times10^{-17}\,{\rm erg\,s}^{-1}\,{\rm cm}^{-2}$. \par 
The fraction of IR AGN detected in the X-rays would be larger in the DS than in the UDS because the DS covers a larger area than the UDS and, therefore, it would contain on average brighter AGN than the UDS. This is visible in Figure \ref{fig:Lbol}, where we show the overall distribution of the bolometric AGN luminosity obtained in the two surveys.  \par 
A second peak is present in the redshift evolution of some fractions mentioned above, particularly in the DS. This second peak is due to the 9.7 $\mu$m absorption feature that biases IR photometric observations to detect the brightest objects. It is however necessary to consider that the predicted depth of this feature is subject of various uncertainties. First, different torus models predict different 9.7 $\mu$m absorption strength, due to differences in the dust distribution and the silicates composition \citep[e.g.][]{Feltre2012,Hatziminaoglou2015}. Second, while some low-metallicity galaxies with a clear 9.7 $\mu$m absorption feature have been observed \citep[e.g.][]{Thuan1999,Houck2004}, some other studies \citep[e.g.][]{Kulkarni2011} have pointed out a possible reduction of the 9.7 $\mu$m absorption strength with decreasing metallicity. If this is the case, we would expect high-$z$ galaxies to have a less pronounced 9.7 $\mu$m absorption feature compared to low-$z$ galaxies, used as template in our simulation. Therefore, SPICA observations around \textit{z}$=$4 would not be biased towards bright IR galaxies, thus reducing the fraction of IR AGN above each X-ray threshold, i.e. smoothing out the secondary peaks visible in Fig. \ref{fig:Xdet_soft}. \par
The fractions of IR-detected AGN above the considered hard X-ray fluxes would have a redshift evolution similar to the fractions of AGN detected in soft X-ray of the same X-ray surveys, although these fractions would be generally lower. These low fractions are due to the hard X-ray depths, which are shallower than the soft X-ray flux depths. In addition, as redshift increases, observed fluxes corresponds to higher rest-frame energies, where the emission falls like a power law with an exponential cutoff. This effect is larger at 2-10 keV than for 0.5-2 keV. \par
We would expect on average less and 1$\%$ of AGN having hard X-ray fluxes above f$_{2-10keV}>1$ and 3$\times10^{-15}\,{\rm erg\,s}^{-1}\,{\rm cm}^{-2}$ and only 5$\%$ above the hard X-ray flux of 2$\times10^{-16}\,{\rm erg\,s}^{-1}\,{\rm cm}^{-2}$, with a maximum of 30$\%$ around \textit{z}$\sim$1.3. In the DS, the average fraction would increase to 1, 3 and 22$\%$ above 3.3$\times10^{-15}$, 1.5$\times10^{-15}$ and 2$\times10^{-16}\,{\rm erg\,s}^{-1}\,{\rm cm}^{-2}$. The fraction of IR-detected AGN above the deepest hard X-ray flux cut, i.e. 2$\times10^{-16}\,{\rm erg\,s}^{-1}\,{\rm cm}^{-2}$, would show a double peak around \textit{z}$\sim$1.3 and \textit{z}$\sim$4.5 with almost no AGN above the hard X-ray flux cut at \textit{z}$>$6, similarly to the soft X-ray AGN fraction in the DS. The detection in both soft and hard X-ray bands will allow for analysing in more detail the spectral type of the AGN, directly fitting the X-ray spectra or trough the analysis of the hardness ratio. \par
In Figures \ref{fig:Xdet_soft} and \ref{fig:Xdet_hard} we also report the expected fraction of AGN with f$_{0.5-2keV}>4\times10^{-17}\,{\rm erg\,s}^{-1}\,{\rm cm}^{-2}$ and f$_{2-10keV}>2\times10^{-16}\,{\rm erg\,s}^{-1}\,{\rm cm}^{-2}$ detected with OST/OSS, in Channel 1 or 2, and with GEP-I, in band 10 or 16. For OST, the average fraction is 25$\%$ (3$\%$) in the OST-Deep survey of 0.5 deg$^{2}$ and 45$\%$ (25$\%$) in the OST-Wide Survey of 20 deg$^{2}$ in the soft (hard) X-rays. The similar area and depth of the SPICA UDS and OST-Deep survey are reflected in similar fractions in both soft and hard X-rays. The average fractions for GEP-I are slightly smaller, 18$\%$ (5$\%$) and 38$\%$ (15$\%$) for the 3 and 30 deg$^{2}$ survey in soft (hard) X-ray, respectively. Even in this case, the fraction of AGN detected in both IR and X-rays generally decreases with increasing redshift.  \par
It is important to highlight that, as mentioned before, the fraction of IR-detected AGN with faint X-ray fluxes is not limited to low-power AGN, but contains also Compton-thick AGN (i.e. obscured AGN with hydrogen column density log$_{10}(N_{H} / {\rm cm}^{-2}) \geq$24) that are challenging to detect at X-rays, but they become bright at IR-wavelengths. For a more detailed analysis on the possible synergies between SPICA or OST and future X-ray telescopes, such as \textsl{Athena}, we refer to the dedicated work of \citet{Barchiesi2021}.

%%%%%%%%%%%%%%%%%%%%%%%%%%%%%%%%%%%%%%%%%%%%%%%%%%%%%%%%%%%%%
\subsection{Detection of IR spectral features}\label{sec:spectra}
SMI-LR spectroscopy could have been performed simultaneously to the photometric SMI observations for both surveys, as the exposure times were long enough to allow for a complete scan of the sky observed in one pointing. Even taking into account that the spectroscopic limits were significantly shallower than the photometric ones, a fraction of galaxies observed in the photometric UDS and DS would have had low-resolution SMI spectra as well. We predicted the capability of SMI-LR to detect some of the main IR spectral features produced by star formation or AGN activity, considering the expected observational depth of the parallel spectroscopic surveys (see Table \ref{tab:surveys}). \par
The detection of multiple lines enables robust redshift estimates and then, depending on the line, the study of different physical properties of the galaxies, e.g. the nature of the ionising sources, gas densities, gas metal abundances. For a detailed analysis of the spectral capability of the other SPICA spectrometer, i.e. SAFARI, we refer to the work by \citet{Spinoglio2021}. The results presented in the next paragraphs are consistent with results from \citet{Spinoglio2021} for the detection of PAH and fine-structure lines, taking into account that, in the aforementioned paper, there is no attempt at separating the AGN and star-forming galaxy populations and the resulting number counts are, consequently, always larger than the ones presented here.\par
As mentioned previously, the current plan for the GEP spectrometer is to perform only pointed observations, therefore we decided to not include any comparison with SPICA SMI-LR. The comparison between SPICA-LR and OST/OSS is instead possible by considering the six OST/OSS channels with their full resolution, i.e. R$=$300, and not with a reduced resolution of R$=$4, as done for the comparison between the photometric surveys. 

%%%%%%%%%%%%%%%%%%%%%%%%%%%%%%%%%%%%%%%%%%%%%%%%%%%%%%%%
\subsubsection{PAH features}\label{sec:PAH}
\begin{figure}
    \centering
    \includegraphics[width=1\linewidth,keepaspectratio]{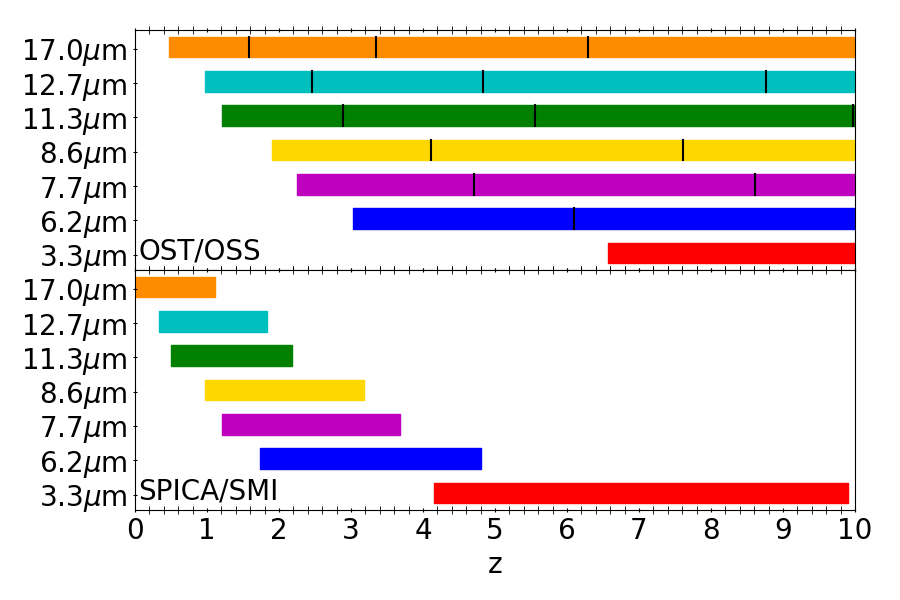}
    \caption{Redshift range in which each PAH feature is inside the wavelength coverage of OST/OSS (\textit{top}) and SPICA/SMI (\textit{bottom}). Vertical black lines show the separation between the different OST/OSS channels. The wide wavelength coverage of OST allows for the simultaneous detection of multiple PAH lines.}
    \label{fig:z_PAH}
\end{figure}

We start our analysis from the PAH features, ranging from 3.3 $\mu$m to 17.0 $\mu$m, derived in \spr{} considering the relations by \citet{Gruppioni2016}, \citet{Bonato2019} and \citet{Mordini2021}. The latter includes different relations for low-metallicty galaxies (i.e. 12+log(O/H)$<$8.2), as some observations point out that these galaxies may have suppressed PAH features \citep{Shipley2016}. In Figure \ref{fig:z_PAH} we report the redshift range in which each PAH feature is inside the wavelength coverage of SPICA/SMI and OST/OSS, for a better comparison between the wavelength coverage of the two instruments. \par

\begin{figure*}[h!]
    \centering
    \includegraphics[width=0.48\linewidth,keepaspectratio]{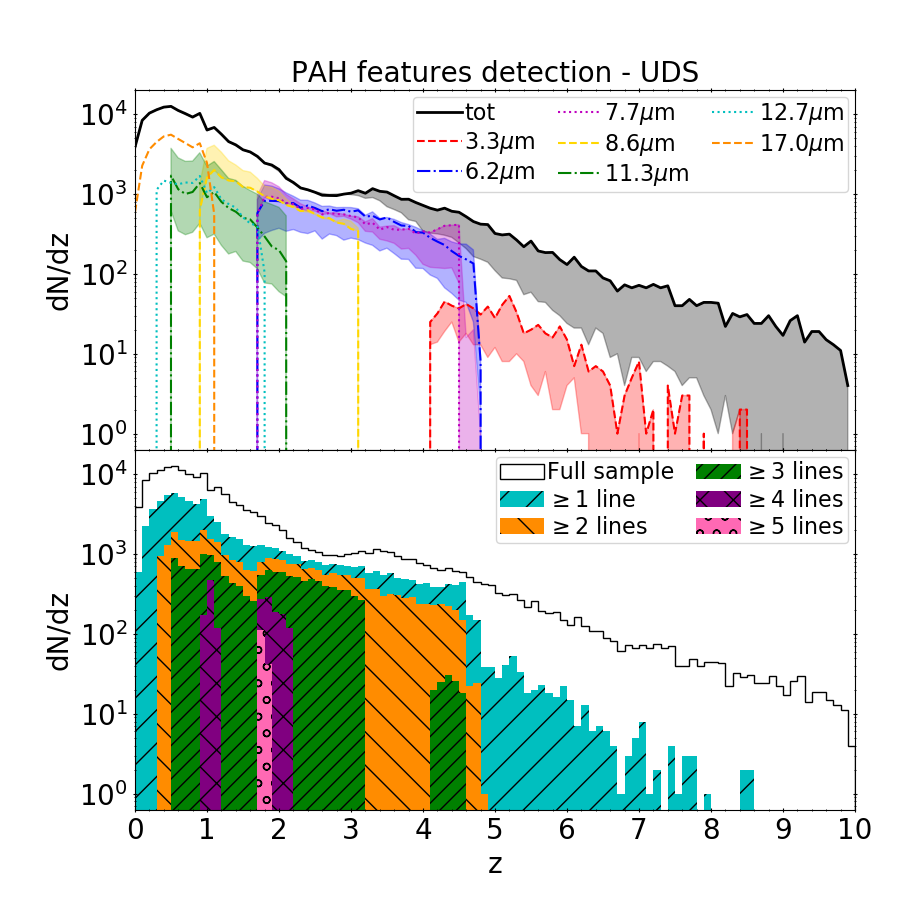}
    \includegraphics[width=0.48\linewidth,keepaspectratio]{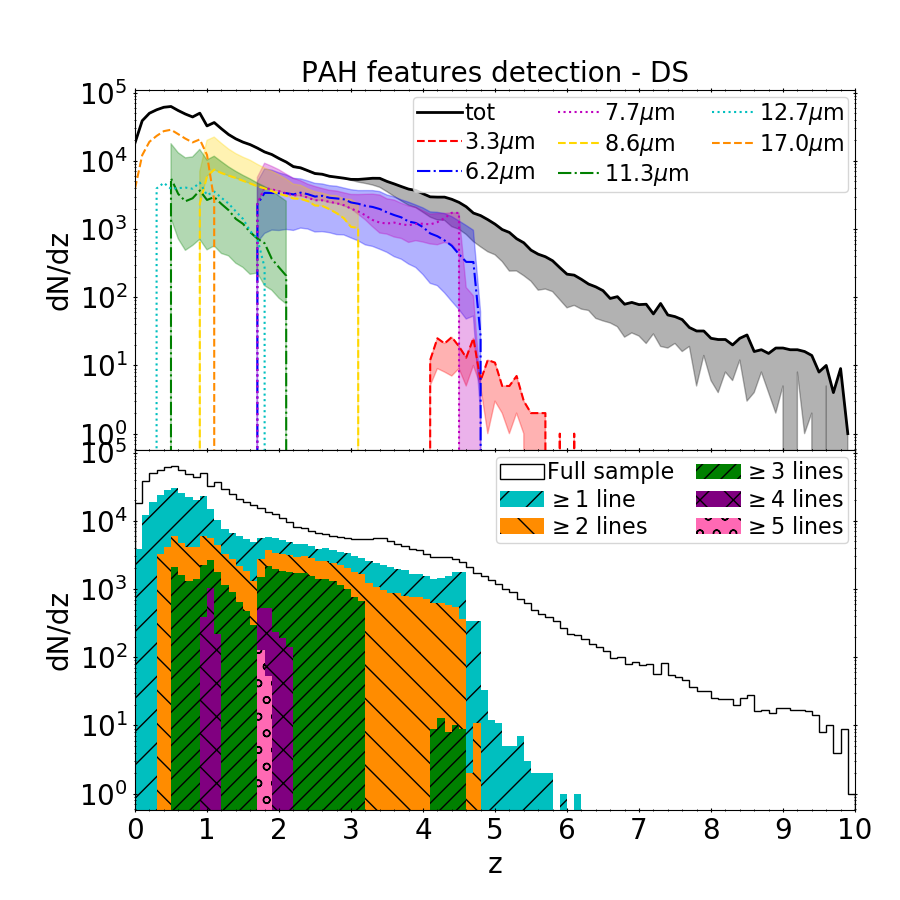}
    \caption{\textit{Top}: Redshift distribution per unit redshift interval of galaxies in the SPICA UDS (\textit{left}) and DS (\textit{right}) with detected PAH lines  (\textit{coloured lines}), i.e. integrated S/N$>$3. We also report the complete sample of galaxies detected with SMI or B-BOP in photometry (\textit{black solid line}). The shaded areas show the uncertainties due to the different empirical relation L$_{IR}$-L$_{line}$, when present, and the extrapolation at \textit{z}$>$3. \textit{Bottom}: Redshift distribution per unit redshift interval of galaxies in the UDS (\textit{left}) and DS (\textit{right}). The sample is divided by the number of detected PAH features. The black solid line is the complete photometric sample. At \textit{z}$<$5, multiple PAH lines would be observed by SPICA in both surveys, for a sizeable amount of galaxies.}
    \label{fig:PAHdet}
\end{figure*}

In Figure \ref{fig:PAHdet} we report, for both the SPICA UDS and DS, the expected redshift distribution per unit redshift interval of galaxies with detections in each PAH feature, where a detection means a S/N$>$3 as derived by integrating the feature over its entire width. In the bottom panels of the same figure we report similar histograms, where we separate galaxies by the number of detected PAHs. This analysis allows us to verify for which and how many galaxies the redshift estimation would be solid, as multiple strong lines would be detectable inside the spectra. \par
In the SPICA UDS, we would expect more than $\sim$17$\%$ of galaxies at \textit{z}$=$1-4.5 to have at least two detected PAH features, with a maximum of $\sim$60$\%$ at \textit{z}$\sim$2.5. These would be mainly star-forming galaxies without an AGN component ($\sim$40$\%$) or composite systems ($\sim$54$\%$). In particular, PAH detections would be extremely rare among dwarf irregular galaxies, i.e. less than 2$\%$ of all dwarf irregular galaxies detected in photometry at \textit{z}$=$1-4.5. The limited number of multiple PAH detections is largely limited by the wavelength coverage of the SPICA/SMI instrument, as seen in Figure \ref{fig:z_PAH}. \par
In the SPICA DS, the fraction varies from 7 to 40$\%$ in the same redshift range, with a maximum at \textit{z}$=$2.6. At \textit{z}$>$5, only the PAH at 3.3 $\mu$m would be covered by the SMI spectral range, thought it would be too faint to be detected for a large fraction (i.e. $>$99$\%$) of the sample. Therefore, for these galaxies the redshift estimation would have to rely on other bright IR emission lines (see Sec. \ref{sec:Hline}-\ref{sec:extraline}) or on the additional information provided by the simultaneous photometric surveys. \par
PAH features are not only important for deriving precise redshifts, but also for identifying potential AGN activity. It has indeed been shown that the equivalent widths of the 6.2 and 11.3 $\mu$m PAH features anti-correlate with the dominance of AGN activity \citep{Spoon2007,Tommasin2008,HernanCaballero2011,Feltre2013}.
%can be used to identify galaxies dominated by AGN activity, as they generally show smaller equivalent widths than galaxies dominated by star formation \citep{Spoon2007,HernanCaballero2011,Feltre2013}. 
This technique to identify potential AGN would be feasible between redshift 0.5 and 5 for 10$^{2}$-10$^{3}$ galaxies per redshift bin (dN/d$z$) in the UDS and DS.\par
\begin{figure*}[h!]
    \centering
    \includegraphics[width=0.48\linewidth,keepaspectratio]{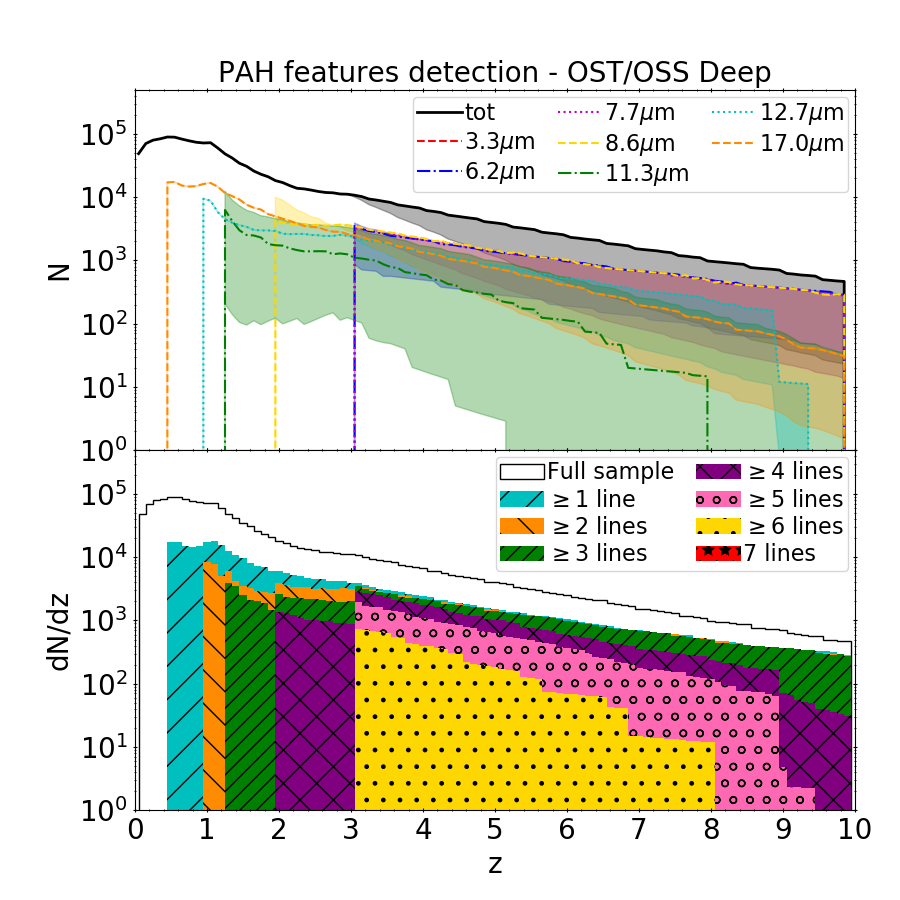}
    \includegraphics[width=0.48\linewidth,keepaspectratio]{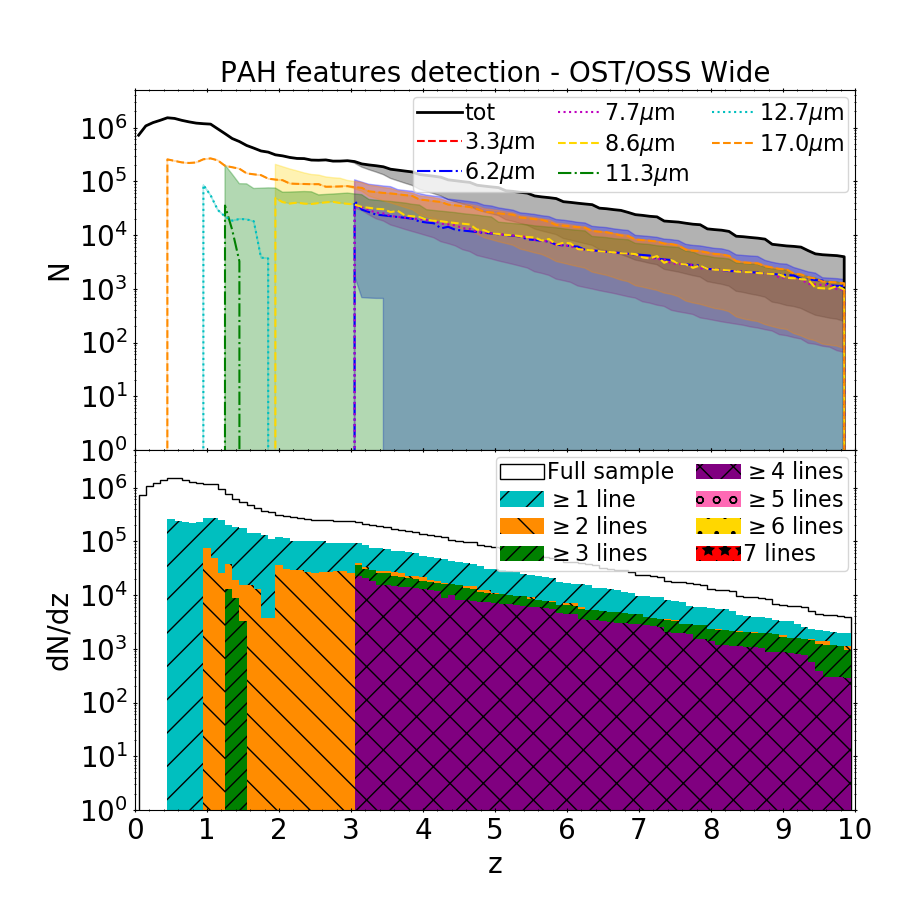}
    \caption{Same as Figure \ref{fig:PAHdet}, but for OST. The complete sample (\textit{black solid line}) consists on galaxies detected in at least one OST/OSS channel considering R$=$4. PAH detections are instead derived considering R$=$300. At \textit{z}$>$1, multiple PAH lines are expected to be observed by OST in both surveys, for a sizeable amount of galaxies.}
    \label{fig:PAHdet_OST}
\end{figure*}

The OST spectroscopic wavelength coverage is larger than the SPICA/SMI one, allowing the simultaneous detection of a significant number of PAH features and over a larger redshift range. This is visible in Figure \ref{fig:PAHdet_OST}, where we show the redshift distribution per unit redshift interval of galaxies with different PAH features detected in at least one of the six OST/OSS channels. \par
In the OST-Deep survey, it will be possible to detect six PAH lines, using the spectroscopic capability of all OST/OSS channels, at \textit{z}$=$3 to 8 for $4\%$ of galaxies on average. The 3.3 $\mu$m PAH feature is expected to be too faint to be detected with OST/OSS, but the application of some wavelength binning can boost the signal, reducing the wavelength resolution, but allowing for the detection of this feature similarly to what is predicted for SPICA/SMI. \par
On average, around 37$\%$ of the galaxies above \textit{z}$=$1 are expected to have at least two PAH features detected. This fraction changes to 11$\%$ in the OST-Wide survey, where we expect to be able to detect at maximum four PAH features at the same time. This happens, on average, for $\sim$7$\%$ of the galaxies at \textit{z}$>$3. The difference is mainly due to the difficulties, given the planned OST-Wide survey observational depths, on detecting the 12.7 and 11.3 $\mu$m features. \par The fraction of galaxies with a detected 11.3 $\mu$m PAH feature is highly uncertain in the OST-Wide survey, depending on the relation considered to derive the flux of this feature, and its detection may be possible at \textit{z}$>$1.5. Even the detection of the 6.2 $\mu$m feature is very uncertain because, when considering the uncertainties in our prediction, it may not be possible to detect this feature at all above \textit{z}$>$3.5. As mentioned for the 3.3 $\mu$m feature, wavelength binning could help on boosting the signal and increase the detection rates for this PAH feature.

\subsubsection{IR hydrogen recombination lines}\label{sec:Hline}
\begin{figure}[h!]
    \centering
    \includegraphics[width=1\linewidth,keepaspectratio]{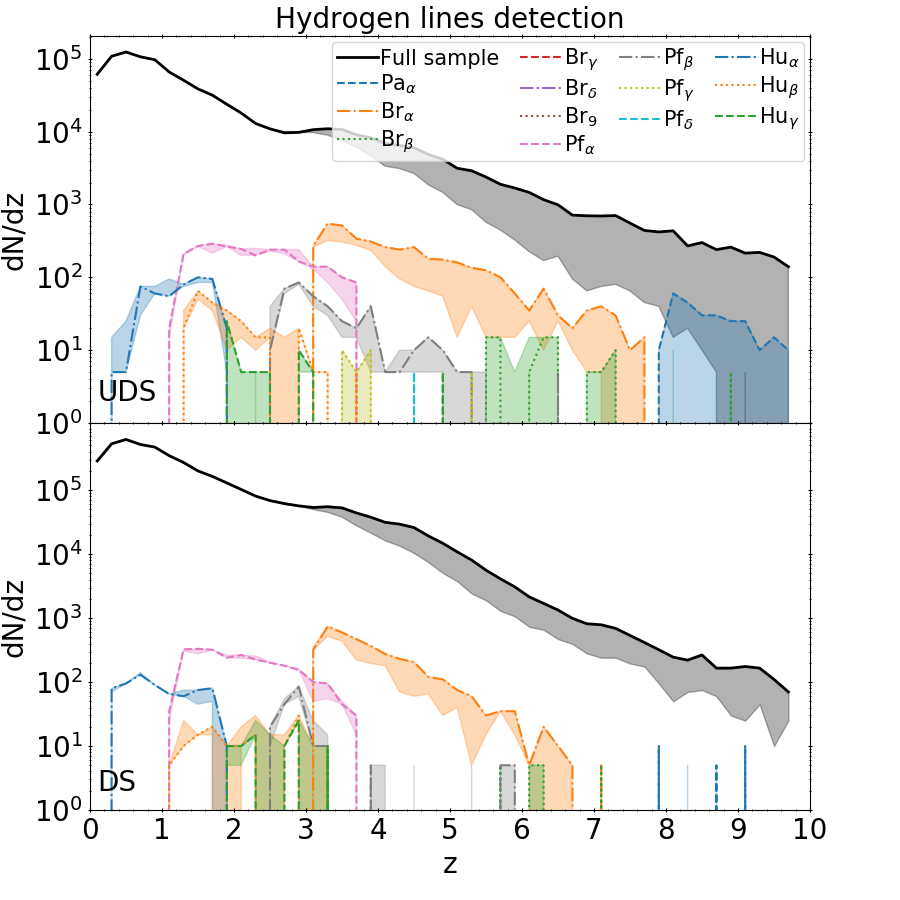}
    \caption{Redshift distribution per unit redshift interval of galaxies with detected (S/N$>$3) hydrogen lines  in the SPICA UDS (\textit{top}) and DS (\textit{bottom}). The list of considered lines is reported in the legend. The shaded areas show the uncertainties due to the extrapolation at \textit{z}$>$3.}
    \label{fig:det_Hline}
\end{figure}

\begin{figure}[h!]
    \centering
    \includegraphics[width=1\linewidth,keepaspectratio]{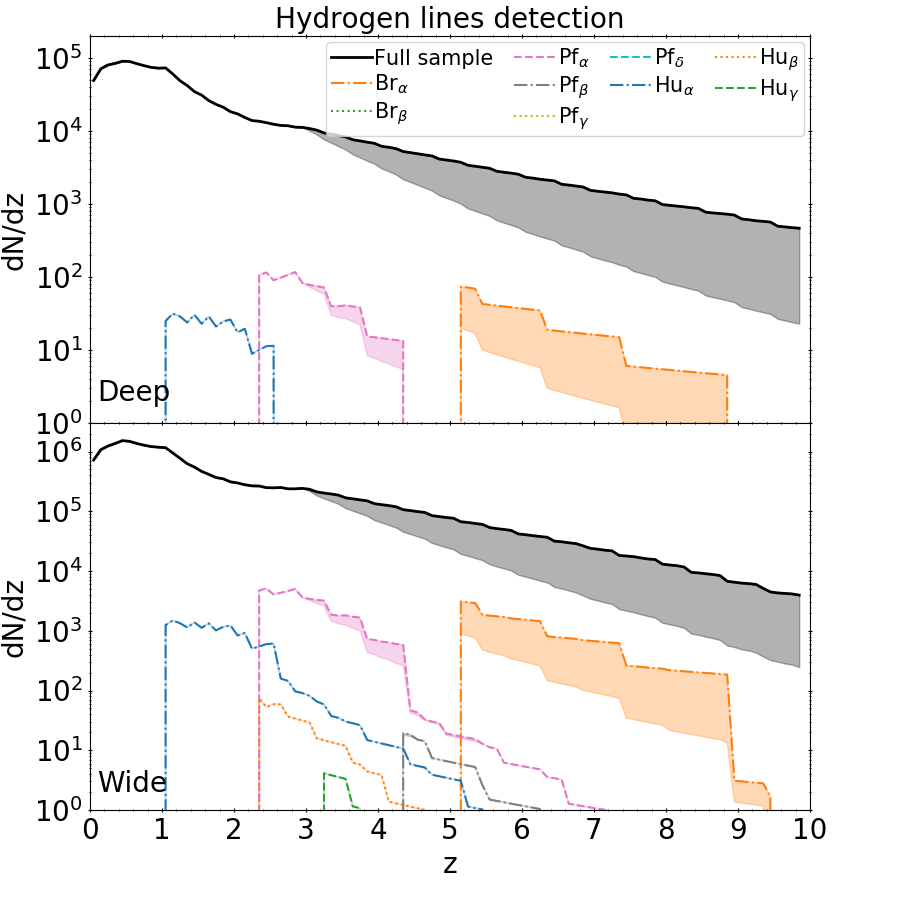}
    \caption{Same as Figure \ref{fig:det_Hline}, but for galaxies detected in at least one OST/OSS channel in the OST-Deep (\textit{top}) and OST-Wide survey (bottom). Only few galaxies are expected to have a detection in an Hydrogen line.}
    \label{fig:det_Hline_OST}
\end{figure}

We now focus on the detection of the main hydrogen nebular emission lines in the IR. In Figures \ref{fig:det_Hline} and \ref{fig:det_Hline_OST} we plot the redshift distribution per unit redshift interval of galaxies with expected detection (S/N$>$3) of different hydrogen recombination lines from the Paschen, Brackett, Pfund and Humphreys series. 
We do not consider hydrogen lines that are outside the SMI or OST wavelength coverage in the considered redshift range. Hydrogen lines include the contribution from both star formation and AGN activity. As the broad-line contribution has not been included in \spr{}, the AGN nebular emission, which accounts only for the emission from the gas in the narrow-line regions, has to be considered as a lower limit to the total (broad+narrow) nebular emission. \par
All the aforementioned hydrogen lines would be too faint to be detected with SPICA/SMI for the large majority of galaxies. In the UDS, only a few percent of galaxies would have detectable Pa$_{\alpha}$, Br$_{\alpha}$, Br$_{\beta}$ or Pf$_{\alpha}$, while all the other lines would be detected for less than 1$\%$ of the total amount of observed galaxies at any redshift. In the DS, only Pa$_{\alpha}$ would be detected for 4$\%$ of the galaxies, while the other hydrogen lines would be detected in less than 1$\%$ of the observed galaxies at any redshifts.\par
In the range between \textit{z}$=$4.9 and \textit{z}$=$8.0, where only the 3.3 $\mu$m PAH feature would be detectable (see Sec. \ref{sec:PAH}), the few detections of the Br$_{\alpha}$ and Br$_{\beta}$ lines could be used for the redshift estimation, for which only the 3.3 $\mu$m PAH feature is otherwise observable. 
At even higher redshifts (\textit{z}$>$8), in the UDS only a few tens of galaxies would have a detection    of the Pa$_{\alpha}$, making the redshift estimation difficult for the rest of the sample. This task would be even more difficult in the DS, where the Pa$_{\alpha}$ line would never be detected. \par
The OST-Deep survey is expected to have a similar depth to the SPICA/SMI UDS survey, but it covers a quarter of the area with a larger wavelength range. The shortest available wavelength is however 25 $\mu$m, compared to 17 $\mu$m of the SPICA/SMI, moving Pa$_{\alpha}$ outside the wavelength coverage at all analysed redshifts. These differences result in the expected detection of few lines, mainly Br$_{\alpha}$, Pf$_{\alpha}$ and Hu$_{\alpha}$, for at maximum 2$\%$ of the galaxies detected in photometry (i.e. OST/OSS with R$=$4), but around 1$\%$ on average. \par
On the OST-Wide survey, the situation is slightly better, given that brighter objects are observed in the OST-Wide survey than in the OST-Deep one. The maximum is reached at \textit{z}$\sim$5, where 4$\%$ of all galaxies observed in photometry have detected Br$_{\alpha}$. In addition, in the OST-Wide survey few galaxies have also detected Hu$_{\beta}$, Hu$_{\gamma}$ and Pf$_{\beta}$, which are expected never to be observed in the Deep survey.\par
Overall, both for both a SPICA-like mission and OST the detection of hydrogen lines is expected to be very challenging and limited to the brightest galaxies. This needs to be taken into account when planning to derive the gas metallicity using the hydrogen lines, although metallicity determinations based only on the IR fine-structure lines have proven to be robust \citep[e.g. Figure 7 in ][]{FernandezOntiveros2021}.

\subsubsection{IR fine-structure lines}\label{sec:extraline}
In this last section we report the predictions for the IR fine-structure lines as included in \spr{},  due to star formation or AGN activity. These fine-structure lines can be used to analyse different physical properties. For example, the use of two lines of the same ionic species, like the two $[\ion{\rm S}{III}]$ lines at 18 and 33 $\mu$m, can be used to investigate the gas density. The use of two lines of the same element, but of two different ionisation levels, like $[\ion{\rm Ne}{III}]$ at 15.55 $\mu$m and $[\ion{\rm Ne}{II}]$ at 12.81 $\mu$m \citep{Thornley2000}, can instead be used to investigate the nature of the ionising source.  In addition, the detection of specific fine-structure lines that need high ionising energies, like the $[\ion{\rm Ne}{V}]$ at 14.32 and 24.31 $\mu$m, is an indication of AGN activity \citep[e.g. ][]{Spinoglio1992,Genzel2000,Melendez2008,Feltre2016}. More details on the use of mid-IR lines to study the nature of ionising sources, the gas densities and the gas metal abundances can be found in \citet{Spinoglio2015} and \citet{FernandezOntiveros2016,FernandezOntiveros2017}.  \par
\begin{figure*}[h!]
    \centering
    \includegraphics[width=1\linewidth,keepaspectratio]{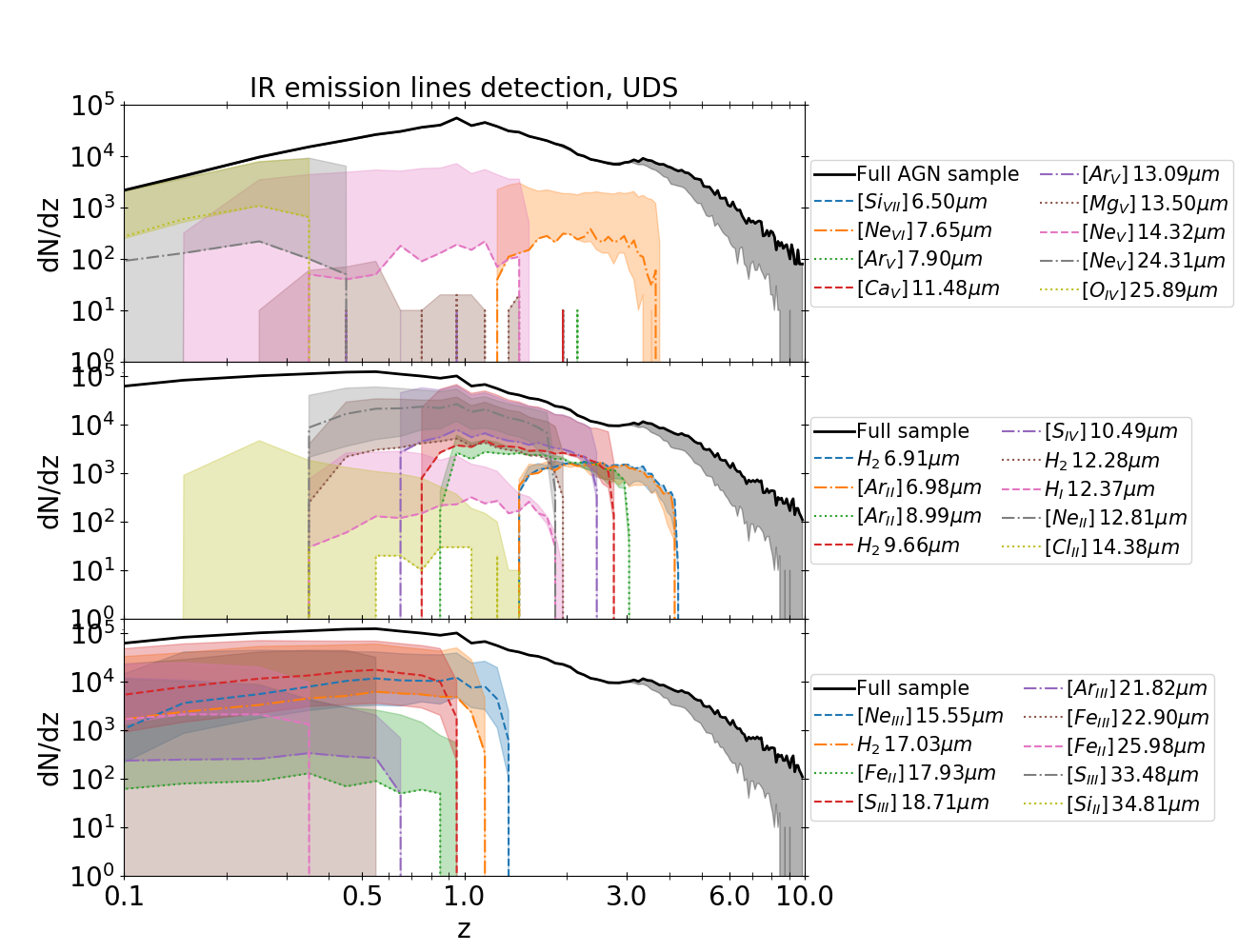}
    \caption{Redshift distribution per unit redshift interval of galaxies in the SPICA UDS with IR fine-structure lines with S/N$>$3. The list of lines is reported on the right of each panel. The first panel show AGN lines while the other two panels show lines originated both by star formation and AGN activity. The complete sample of galaxies detected with SMI or B-BOP in photometry is shown in black, in the top panel we limit the sample to AGN only. We consider as reference the line luminosity derived from the L$_{IR}$ considering the relation by \citet{Bonato2019}. The shaded areas show the uncertainties due to different empirical relations L$_{IR}$-L$_{line}$ \citep[i.e. ][]{Gruppioni2016,Mordini2021}, when present, and the extrapolation at \textit{z}$>$3.}
    \label{fig:Linedet_UDS}
\end{figure*}
\begin{figure*}[h!]
    \centering
    \includegraphics[width=1\linewidth,keepaspectratio]{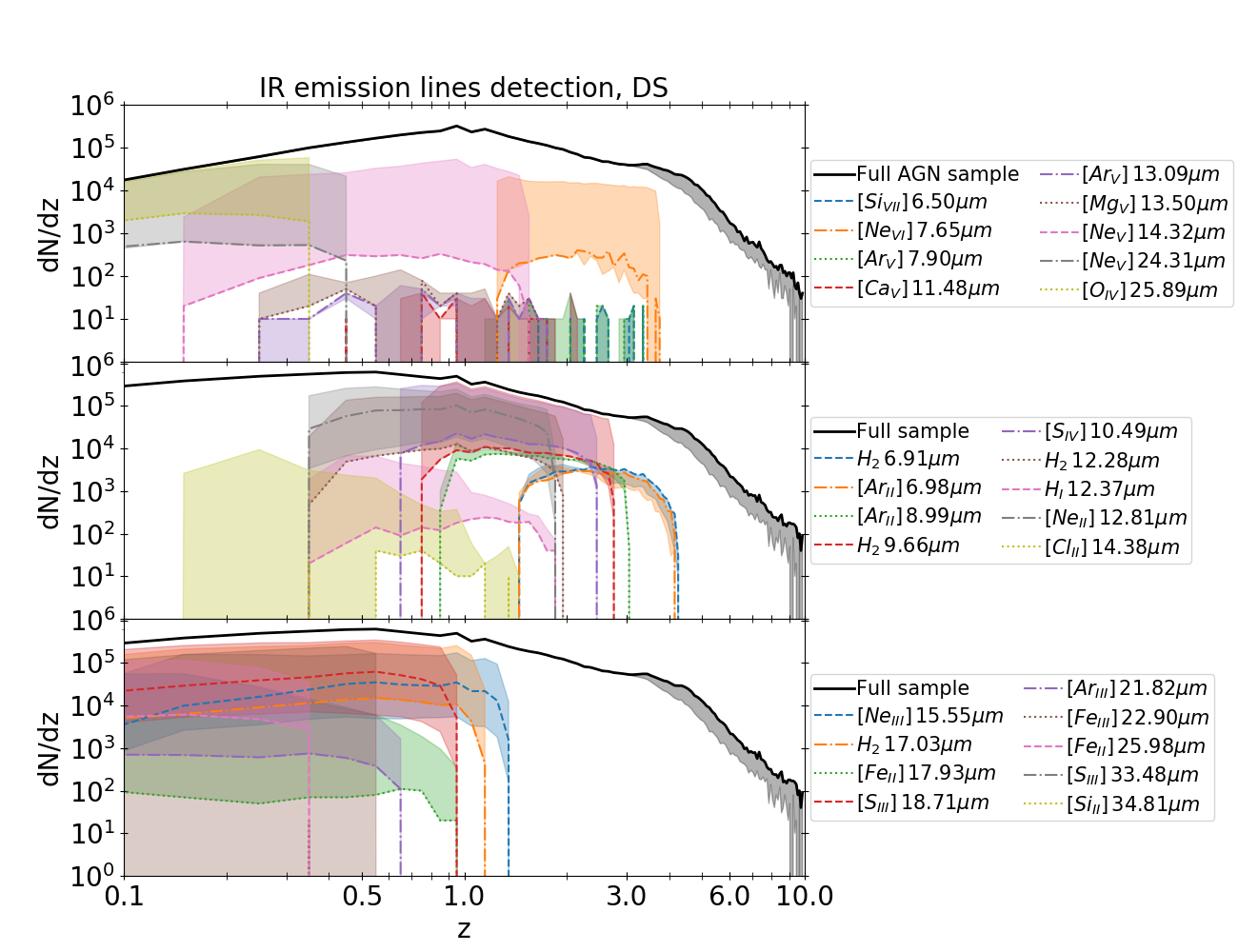}
    \caption{Same as of Figure \ref{fig:Linedet_UDS}, but for the SPICA DS.}
    \label{fig:Linedet_DS}
\end{figure*}

The redshift at which we would expect each fine-structure line to be observed is shown in Figures \ref{fig:Linedet_UDS} and \ref{fig:Linedet_DS}, for the SPICA UDS and DS, respectively.  The detection of these lines with SMI-LR would be limited at \textit{z}$<$4.5, because of the SMI wavelength coverage. At higher redshift, it would be necessary to search for emission lines in the rest-frame near-IR, precisely in the range 1.5 $\mu$m $<\lambda<$ 6.5 $\mu$m that are not currently included in \spr{}. \par
There would be no significant difference in the relative numbers of detections between the UDS and DS. The precise percentage of detections depends on the considered lines and redshift, ranging from less than 1$\%$ to almost all the galaxies. This gives an idea of the great power of such IR spectroscopy, offering a plethora of nebular emission lines not only to derive a precise redshift estimation, but also to study both star formation and AGN activity.\par
\begin{figure*}[h!]
    \centering
    \includegraphics[width=1\linewidth,keepaspectratio]{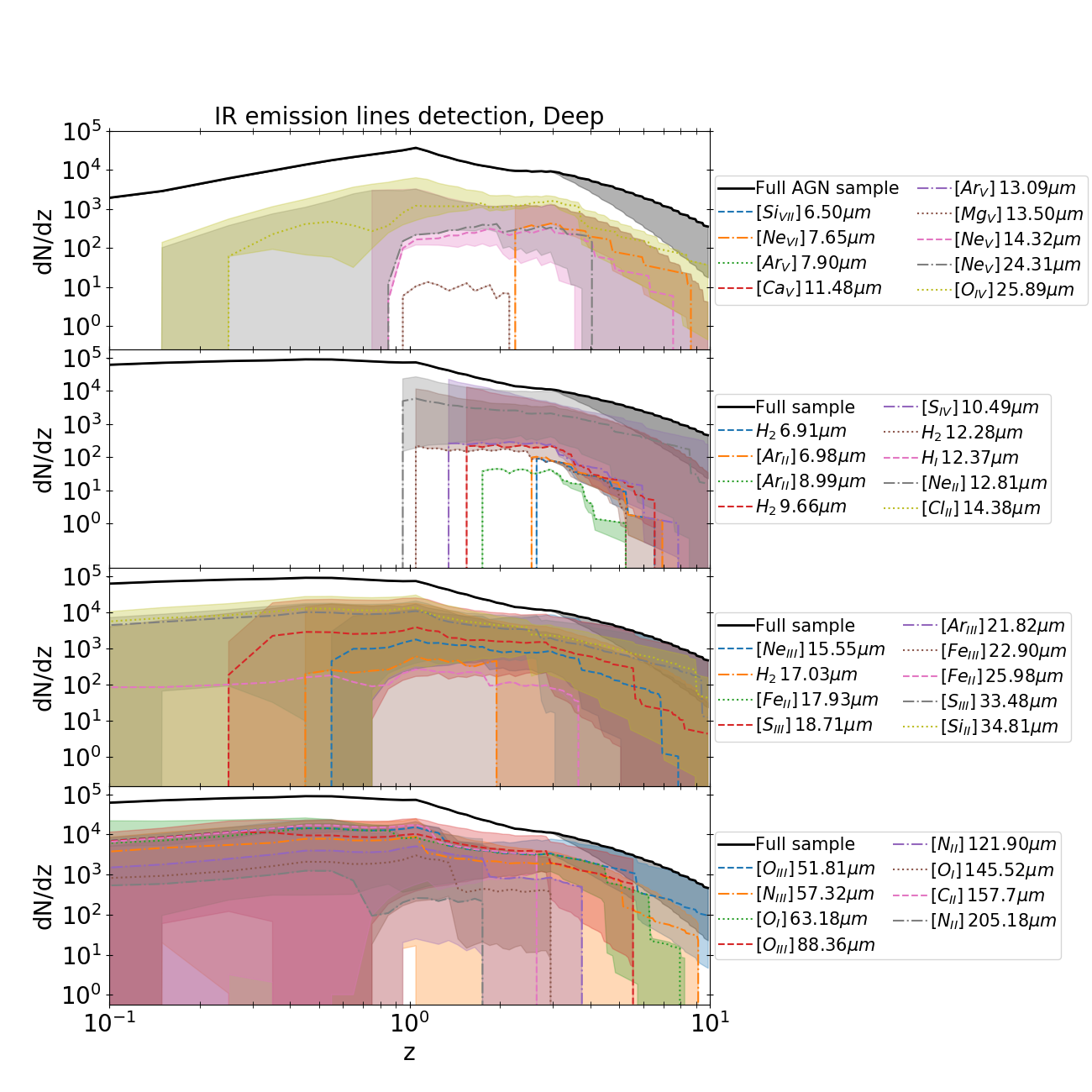}
    \caption{Same as of Figure \ref{fig:Linedet_UDS}, but for the OST-Deep survey. The complete sample refers to galaxies detected in at least one OST/OSS channel considering R$=$4, while line detections are instead derived considering R$=$300.}
    \label{fig:Linedet_DOST}
\end{figure*}
\begin{figure*}[h!]
    \centering
    \includegraphics[width=1\linewidth,keepaspectratio]{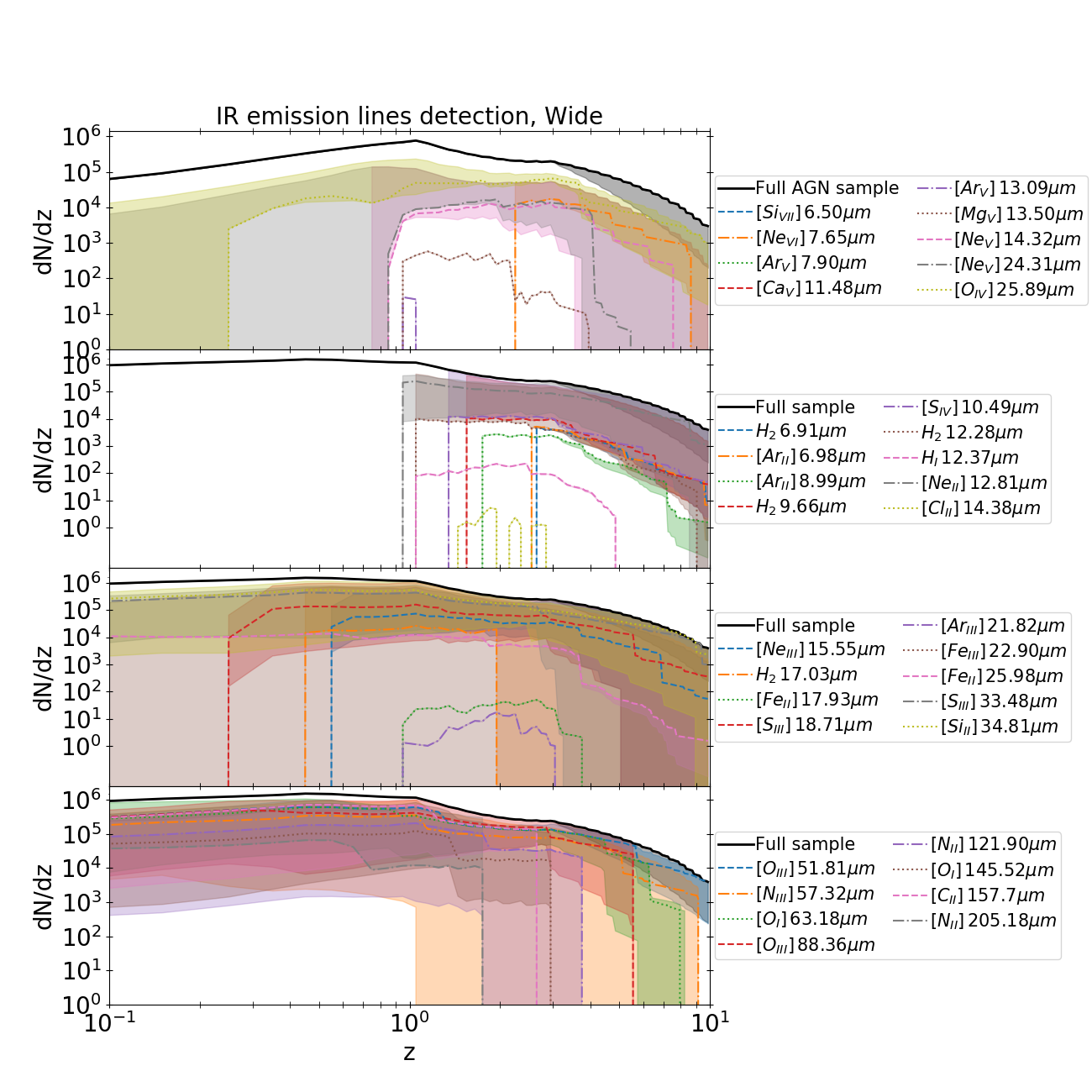}
    \caption{Same as Figure \ref{fig:Linedet_UDS}, but for the OST-Wide survey. The complete sample refers to galaxies detected in at least one OST/OSS channel considering R$=$4, while line detections are instead derived considering R$=$300.}
    \label{fig:Linedet_WOST}
\end{figure*}

In Figures \ref{fig:Linedet_DOST} and \ref{fig:Linedet_WOST}, we report results for the OST/OSS, considering all six available channels. We consider also additional fine-structure lines at wavelengths longer than what would be observed by SPICA/SMI, i.e. $\lambda>$37 $\mu$m, as OST/OSS covers up to 589 $\mu$m. The number of galaxies with fine-structure-line detections varies with redshift, line and survey.\par
In the OST-Deep survey this number is generally below 10$^{4}$ galaxies per redshift bin, while in the OST-Wide surveys some lines, like the $[\ion{\rm O}{I}]$ at 63 $\mu$m, are expected to be detected in 5$\times10^{5}$ galaxies per redshift bin, at least at very low-$z$ (i.e. \textit{z}$<$0.8). The redshift range in which the different lines can be detected is very large, compared to what would have been possible with SPICA, thanks to the OST/OSS large wavelength range. Therefore, we expect that the analysis of many fine-structure lines, such as the $[\ion{\rm Ne}{V}]$ at 14.32 and 24.3 $\mu$m and $[\ion{\rm O}{IV}]$ at 25.9 $\mu$m, tracers of AGN activity, can be extended even at \textit{z}$>$4.5 with future OST/OSS observations.

% \begin{table}[]
%     \centering
%     \begin{tabular}{c|cc|cc}
%          Line & \multicolumn{4}{c}{Number of detections} \\
%           & \multicolumn{2}{c|}{SPICA} & \multicolumn{2}{c}{OST} \\
%           & UDS & DS & Deep & Wide \\
%           \hline
%          $[\ion{\rm Si}{VII}]$ 6.50$\mu$m & 16 & & & \\ 
%          H$_{2}$ 6.91$\mu$m & 2600-6400 & & & \\
%          $[\ion{\rm Ar}{II}$ 6.98$\mu$m & 1900-5400 & & & \\
%          $[\ion{\rm Ne}{VI}]$ 7.63 $\mu$m & 160-6.9$\times$10$^{4}$ & & & \\
%          $[\ion{\rm Ar}{V}]$ 7.90$\mu$m & 20 & & & \\
%          $[\ion{\rm Ar}{III}$ 8.99$\mu$m &  & & & \\
%          $[\ion{\rm Ca}{V}]$ 11.48$\mu$m & 16 & & & \\ 
%          $[\ion{\rm Ar}{V}]$ 13.09$\mu$m & 1-19 & & & \\ 
%          $[\ion{\rm Mg}{V}]$ 13.50$\mu$m & 9-29 & & & \\ 
%          $[\ion{\rm Ne}{V}]$ 14.32$\mu$m & 267-1.4$\times$10$^{5}$ & & & \\
%          $[\ion{\rm Ne}{V}]$ 24.31$\mu$m & 195-2.2$\times$10$^{4}$ & & & \\
%          $[\ion{\rm O}{IV}]$ 25.89$\mu$m & 72-1.1$\times$10$^{4}$ & & & \\
         
%     \end{tabular}
%     \caption{Number of fine-structure lines that would be detected by SPICA/SMI and that will be detected by OST/OSS, considering the entire redshift range analysed in the paper. When the lines are derived considering different line-luminosity relations, we report the minimum and maximum values among the different estimations.}
%     \label{tab:finestructure_lines}
% \end{table}
%%%%%%%%%%%%%%%%%%%%%%%%%%%%%%%%%%%%%%%%%%%%%%%%%%%%%%%%%%%%%%%%%%%%%%%%%%%%%%%%%%%%%%%%%%%%%%%%%

\section{Conclusions}\label{sec:conclusions}
In this paper we presented a set of predictions for the spectro-photometric surveys planned for the SPICA mission concept, derived with the state-of-the-art simulation \spr{}. Results are valid also for similar 2.5 m actively cooled space missions. We also present detailed predictions for NASA's OST and GEP mission concepts, obtained with the same simulation. \par
In particular, we simulated the SPICA-like DS and UDS considering galaxies with detection in at least one of the broad-band filters centred at 34 (SMI) and 70 $\mu$m (B-BOP) and observations with the simultaneous SMI-LR spectrometer ranging from 17 to 36 $\mu$m. We considered a 5$\sigma$ depth of 3 and 13 $\mu$Jy at 34 $\mu$m and 60 and 100 $\mu$Jy at 70 $\mu$m, in the UDS and DS respectively. We also simulated two surveys for OST, a Deep one of 0.5 deg$^{2}$ and a Wide one of 20 deg$^{2}$, and other two surveys of 3 and 30 deg$^{2}$ for GEP.\par
Our main findings are:
\begin{itemize}
    \item The most luminous galaxies would be observed up to \textit{z}$=$10, in the SPICA DS of 15 deg$^{2}$, and, even considering the most conservative results, at least up to \textit{z}$=$8 in the SPICA UDS. Similar conclusions apply also to OST and GEP.
    \item LIRGs would be observed at 34 $\mu$m up to \textit{z}$=$4.0 in the UDS and up to \textit{z}$=$2.5 in the DS, whereas ULIRGs would be detected at all redshifts in the both the UDS and DS. Observations of LIRGs would be mainly limited at \textit{z}$<$2 at 70 $\mu$m in both surveys. The observational depth planned for the OST/OSS Chanel 2 in the OST-Deep survey should instead allow for detecting at least few LIRGS up to \textit{z}$=$4.
    \item On average, 26$\%$ (37$\%$) of AGN detected at 34 or 70 $\mu$m in the SPICA UDS (DS) would have soft 0.5-2 keV X-ray fluxes above 4$\times$10$^{-17}\,{\rm erg\,s}^{-1}\,{\rm cm}^{-2}$, i.e. corresponding to the observational depth of the  \textsl{Athena} Deep Survey. The fractions are lower if we consider detections in the hard X-rays. This shows the complementary nature of X-ray and IR observations in the study of AGN in general, and obscured AGN in particular.
    \item The redshift estimation could be derived for SPICA/SMI using at least two PAH features from \textit{z}$=$1.0 to \textit{z}$=$5, for 29$\%$ (19$\%$) of the sample in the UDS (DS) on average. At \textit{z}$\sim$5-8, the redshift estimation could be based on the Br$_{\alpha}$ and Br$_{\beta}$ lines at these redshifts, at least for the few galaxies for which these lines are detected. At higher redshifts, only few tens of galaxies could have a detection in an hydrogen recombination line. 
    \item Considering all OST/OSS channels and galaxies detected in photometry (i.e. R$=$4) in the OST-Deep survey, we expect to detect six different PAH features for 4$\%$ of the galaxies at \textit{z}$=$3-8 and at least two PAH features for 37$\%$, on average, of galaxies at \textit{z}$>$1. However, OST wavelength range, starting at wavelengths longer than SPICA/SMI, limits the number of observable bright hydrogen recombination lines. 
    \item The AGN and host galaxy analysis could be performed with SPICA/SMI at \textit{z}$<$4.5 and it will be possible at even higher redshfits with OST/OSS. This is possible using a large set of IR nebular emission lines originated from star formation or AGN activity, such as different neon lines like the $[\ion{\rm Ne}{V}]$ at 14.32 $\mu$m and at 24.31 $\mu$m and the $[\ion{\rm Ne}{II}]$ at 12.81 $\mu$m.
\end{itemize}
These results show the capability of SPICA, OST and GEP to detect galaxies up to \textit{z}$=$10 without being affected by dust obscuration and also to study the physics of AGN and star formation for a large fraction of such objects.

\begin{acknowledgements}
This paper is dedicated to the memory of Bruce Swinyard, who initiated the SPICA project in Europe, but unfortunately died on 22 May 2015 at the age of 52. He was ISO-LWS calibration scientist, \hers-SPIRE instrument scientist, first European PI of SPICA and first design lead of SAFARI. We acknowledge the whole SPICA Collaboration Team.
LB, CG, LS and JAFO acknowledge financial support by the Agenzia Spaziale Italiana (ASI) under the research contract 2018-31-HH.0. LB would like to thank F. La Franca and M. Malkan for providing constructive comments. AF and FC acknowledges the support from grant PRIN MIUR2017-20173ML3WW\_001.
MPS acknowledges support from the Comunidad de Madrid through the Atracci\'on de Talento Investigador Grant 2018-T1/TIC-11035 and PID2019-105423GA-I00 (MCIU/AEI/FEDER,UE). FJC acknowledges financial support from the Spanish Ministry MCIU under project RTI2018-096686-B-C21 (MCIU/AEI/FEDER/UE), cofunded by FEDER funds and from the Agencia Estatal de Investigación, Unidad de Excelencia María de Maeztu, ref. MDM-2017-0765.
\end{acknowledgements}
%% Aggiungere negli aknoledgment Fabio La Franca per aver inviato i commenti

% \begin{appendix}

% \section{AN EXAMPLE OF APPENDIX HEAD}

% \end{appendix}

\bibliographystyle{pasa-mnras}
\bibliography{mybib}

\end{document}